\documentclass[aps,twocolumn,prd,superscriptaddress,showpacs,preprintnumbers,nofootinbib]{revtex4-1}

\usepackage{mathrsfs}

\usepackage[utf8]{inputenc}
\usepackage{dsfont}
\usepackage{graphicx}
\usepackage{slashed}
\usepackage{amsmath,amssymb,amsthm,amsfonts}
\usepackage{dsfont,bbm}
\usepackage{color,rotating}
\usepackage{units}
\usepackage{appendix}
\usepackage{hyperref}
\usepackage{subcaption}
\usepackage{xcolor}

\def\Fig#1{Fig.~\ref{#1}}
\def\Tab#1{Tab.~\ref{#1}}

\def\Eq#1{Eq.~(\ref{#1})}
\def\eq#1{(\ref{#1})}
\def\App#1{Appendix~\ref{#1}}
\def\Sec#1{Section.~\ref{#1}}

\newcommand{\imag}{\text{i}}

\newcommand{\sign}{{\text{sign}}}

\graphicspath{{./figures/}}

\def\eq#1{(\ref{#1})}

\newcommand{\gettitle}{Finite temperature spectral functions in the $O(N)$-model}

\hypersetup{
	colorlinks,
	linkcolor={red!75!black},
	citecolor={blue!75!black},
	urlcolor={blue!75!black},
	pdftitle={\gettitle},
	pdfauthor={Pawlowski, Strodthoff, Wink},
	pdfkeywords={Spectral function} {Analytic continuation}
	{correlations functions} {temperature} 
	{functional renormalisation group}
	{real time} {order parameter},
	bookmarksopen=true,
	bookmarksopenlevel=2,
	bookmarksnumbered=true
}

\begin{document}
	
\title{\gettitle}
	
\author{Jan M. Pawlowski}
\affiliation{Institut f\"ur Theoretische Physik, Universit\"at
  Heidelberg, Philosophenweg 16, 69120 Heidelberg, Germany}
\affiliation{ExtreMe Matter Institute EMMI, GSI, Planckstr. 1, D-64291
  Darmstadt, Germany}
	
\author{Nils Strodthoff}
\affiliation{Nuclear Science Division, Lawrence Berkeley National
  Laboratory, Berkeley, CA 94720, USA}
	
\author{Nicolas Wink} \affiliation{Institut f\"ur Theoretische Physik,
  Universit\"at Heidelberg, Philosophenweg 16, 69120 Heidelberg,
  Germany}

\begin{abstract}
	
We directly calculate spectral functions in the $O(N)$-model at
finite temperature within the framework of the Functional
Renormalization group. Special emphasis is put on a fully numerical
framework involving four-dimensional regulators preserving Euclidean
$O(4)$ and Minkowski Lorentz invariance, an important prerequisite
for future applications. Pion and sigma meson spectral functions are
calculated for a wide range of temperatures across the phase
transition illustrating the applicability of the general framework
for finite temperature applications. In addition, various aspects
concerning the interplay between the Euclidean and real time
two-point function are discussed.

\end{abstract}
	
\maketitle

\section{Introduction}
	
The access to real time correlation functions is key to the
theoretical understanding of many interesting physics phenomena,
ranging from the bound state spectrum of the theory at hand over
decays to the dynamical evolution of both, close-to- and
far-from-equilibrium systems. However, strongly correlated systems are
typically only accessible within numerical computations with either
lattice or functional approaches. These approaches have been mostly
used in their Wick rotated form in Euclidean space if it comes to
applications to strongly correlated systems in higher dimensions such
as QCD.

In the present work we report on progress within a fully numerical
approach to directly compute real time correlation functions,
introduced in~\cite{Pawlowski:2015mia,Strodthoff:2016pxx}.  This
formalism is based on the functional renormalisation group
(FRG)~\cite{Wetterich:1992yh}, a suitable non-perturbative method, for
reviews see e.g.\ \cite{Berges:2000ew,Pawlowski:2005xe}.  Such a
numerical framework is required for application to strongly-correlated
systems whose degrees of freedom have a non-trivial dispersion
relation. Furthermore, already in Euclidean applications the use of
regulators preserving Euclidean $O(4)$ and Minkowski Lorentz
invariance is crucial in situations with finite temperatures and
density. We would also like to emphasise that the current real time
approach is useful and applicable to finite density situations in
Euclidean space: the $O(4)$-invariant regulators and methods discussed
here allow us to guarantee the silver-blaze property without violating
Lorentz symmetry. Moreover, the flow is local in both momentum- and
frequency space. It is this combination of properties which is crucial
for a quantitative determination of the density. The above
requirements are met in QCD at finite temperature and density but also
in ultracold atomic systems. Both systems are key application areas of
the current real time approach.

FRG approaches to real time computations close to the present one have
been set-up in \cite{Floerchinger:2011sc, Kamikado:2013sia}. While the
former one shares the space-time symmetry with the present approach,
the latter one shares the locality in momentum space. The present
approach combines both momentum- and frequency locality as well as
Lorentz invariance.

Real-time applications range from the dynamics of low dimensional
systems to the description of far-from-equilibrium dynamics, for a
selection of works see \cite{jakobs2007nonequilibrium,gezzi2007functional,
    Gasenzer:2008zz,karrasch2008finite,Pietroni:2008jx,
	Berges:2008sr,Sinner:2009zz,Canet:2009vz,bartosch2009functional,
	Gasenzer:2010rq,karrasch2010functional,karrasch2010finite,jakobs2010nonequilibrium,
	Serreau:2011fu,
	kashuba2013quench,Mathey:2014xxa,2014PhRvB..89m4310S,Mesterhazy:2013naa,Floerchinger:2016hja,
	Prokopec:2017vxx,Buessen:2017qui}
Most
of the implementations have in common that either they are not
directly applicable to higher dimensional systems or one would have to
re-derive the plethora of Euclidean results already obtained in the
Euclidean version of the approach. This situation asks for an approach
which allows us to either make direct use of available Euclidean results 
or at least be able to benchmark the real time results with the Euclidean
ones already obtained. In addition, the physical content of Euclidean truncations
in terms of scattering processes, resonances, particles and decay channels can be
directly accessed from the real time correlation functions and identified with the
equivalent Euclidean part.  

In principle it is possible to solve the hard problem of obtaining
real time correlation functions from reconstruction of Euclidean
data. This tasks requires a very good resolution of the momentum and
frequency dependence of the Euclidean correlation functions, for
related FRG-work see \cite{Ellwanger:1995qf,Bergerhoff:1997cv,2009PhRvB..79s5125H,
  Blaizot:2005xy,Blaizot:2005wd,Blaizot:2006vr,Benitez:2011xx,
  Mitter:2014wpa,Christiansen:2015rva,Cyrol:2016tym,Cyrol:2017ewj,Cyrol:2017qkl,Denz:2016qks}.
Subsequently applied methods based on Bayesian Reconstruction, like
the Maximum Entropy Method (MEM) \cite{Jarrell:1996rrw,
  Asakawa:2000tr, Engels:2009tv, Pawlowski:2016eck, Rothkopf:2016luz,
  Ilgenfritz:2017kkp} are able to produce remarkable results in many
cases. For spectral reconstrunctions based on Euclidean FRG data see
e.g.\
\cite{Schmidt:2011zu,Haas:2013hpa,Christiansen:2014ypa,Rose:2015bma}.
Unfortunately, these methods are mathematically not guaranteed to
converge towards the correct correlation function. Other methods which do not have this problem
\cite{Cuniberti:2001hm, Burnier:2011jq, Ferrari:2016snh}, suffer from
sign problems and require unachievable numerical precision for
reliable results. All of these obstacles can be overcome by a direct
numerical calculation.

Our long term goals are the hadron spectrum, transport coefficients
and other real time observables within the framework of the fQCD
collaboration~\cite{fQCD:2016-10}, which aims at a quantitative
first-principle description of QCD from the FRG. This stresses once
more the importance of a fully numerical approach due to the
sophisticated technical level already required at the Euclidean level
to obtain quantitatively competitive results~\cite{Mitter:2014wpa,
Cyrol:2016tym, Cyrol:2017ewj, Cyrol:2017qkl}. This work is a
necessary piece in order to close the gap between the calculation
of Euclidean correlation functions and dynamical observables
such as transport coefficients.
These quantities can be obtained conveniently from the spectral functions
as demonstrated with the shear viscosity~\cite{Haas:2013hpa,Christiansen:2014ypa}
from reconstructed spectral functions.

\section{$O(N)$ model spectral functions at finite temperature} \label{sec:formalism} 
The calculation of real time
observables such as spectral functions starting from a Euclidean
formalism represents a difficult problem as it requires the analytic
continuation from Euclidean to Minkowski signature.  In this work we
follow the formalism put forward in \cite{Pawlowski:2015mia} where the
analytic continuation is carried out on the level of the
equation itself.  Unfortunately the straightforward evaluation of
Euclidean correlation function for complex momenta does not lead to
the desired retarded correlation functions required for the extraction
of the spectral function. The procedure in \cite{Pawlowski:2015mia}
thus requires a deformation of the integration contour into the
complex plane with appropriate corrections in order to recover the
desired real time observable. The following discussion is restricted
to the simplified case, referred to as propagator approximation in the
following, where only the Minkowski-momentum dependence of propagators
is taken into account. This approximation obviously misses bound-state
effects in $n-$point vertices that are hidden in a non-trivial
Minkowski-momentum dependence of the vertices. However, such effects
could still be captured by a dynamical hadronisation of the
corresponding channels in the respective vertices within this
approximation.
	
\begin{figure}
  \includegraphics[width=.7\linewidth]{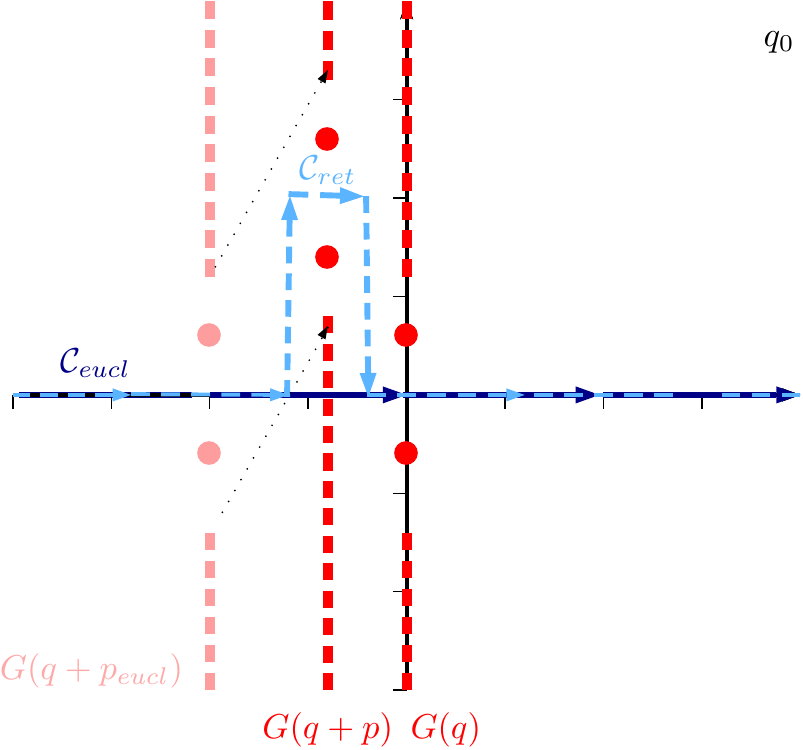}
  \caption{Integration path in the complex $q_0$ plane at vanishing
    temperature}
  \label{fig:continuationT0}
\end{figure}
Here we present a formal solution which holds in the propagator
approximation irrespective of the analytic structure of the
propagators under consideration. Following \cite{Pawlowski:2015mia},
we start by recapitulating the case of vanishing temperature. In this
approximation it is sufficient to consider a single diagram topology,
namely the self-energy contribution to the two-point function
involving two three-point vertices. Here we are interested in
contributions to the retarded 2-point functions as the latter relates
directly to the spectral function. Given analytical propagators as a
function of a complex momentum variable, it turns out that the
contribution to the retarded correlation function is obtained by
integration along a particular path in the complex momentum plane. The
defining property of this path is that it represents a continuous
deformation of the integration path along the real axis for vanishing
Minkowski external momentum. This is illustrated in
Fig.~\ref{fig:continuationT0} for the case of two propagators which
only carry a complex structures on the Minkowski axis, but this
argument generalizes straightforwardly to propagators with a general
complex structure. In the case of only simple poles the contribution
arising from integrating along the simple contour $\mathcal{C}_{eucl}$
relates to the desired contribution to the retarded correlator
obtained by integrating along $\mathcal{C}_{ret}$ by a closed contour
integration. The difference between the two results can be written as
a sum of residues of certain poles in the complex plane, which is
exactly the situation covered in \cite{Pawlowski:2015mia}.
	
\begin{figure}
		\includegraphics[width=.7\linewidth]{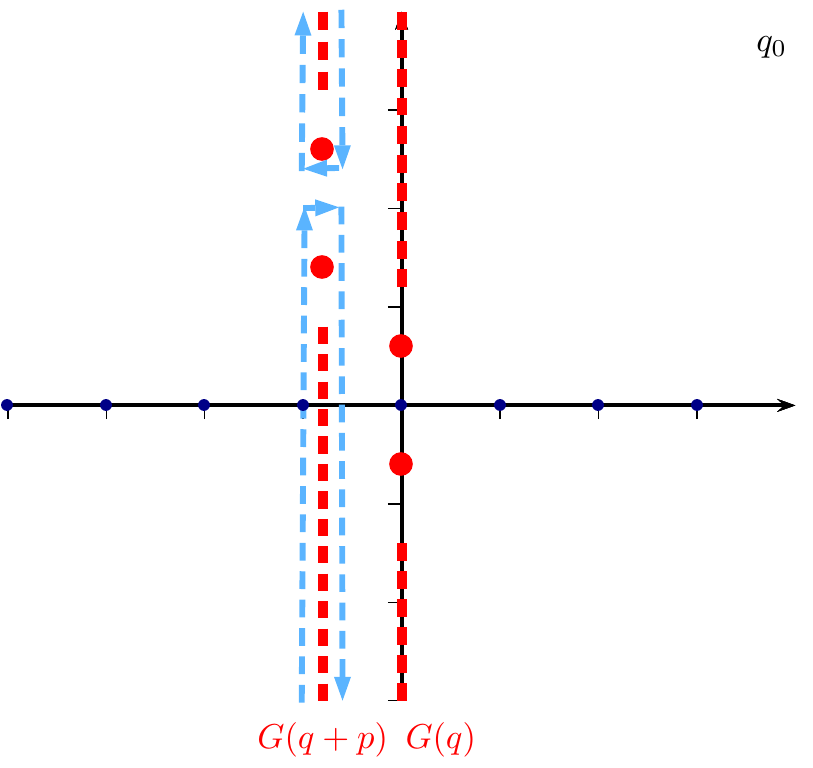}
		\caption{Integration path in the complex $q_0$ plane
                  at finite temperature}
		\label{fig:continuationT}
\end{figure}
\begin{figure*}
	\begin{tabular}{cc}
		\includegraphics[width=0.5\linewidth]{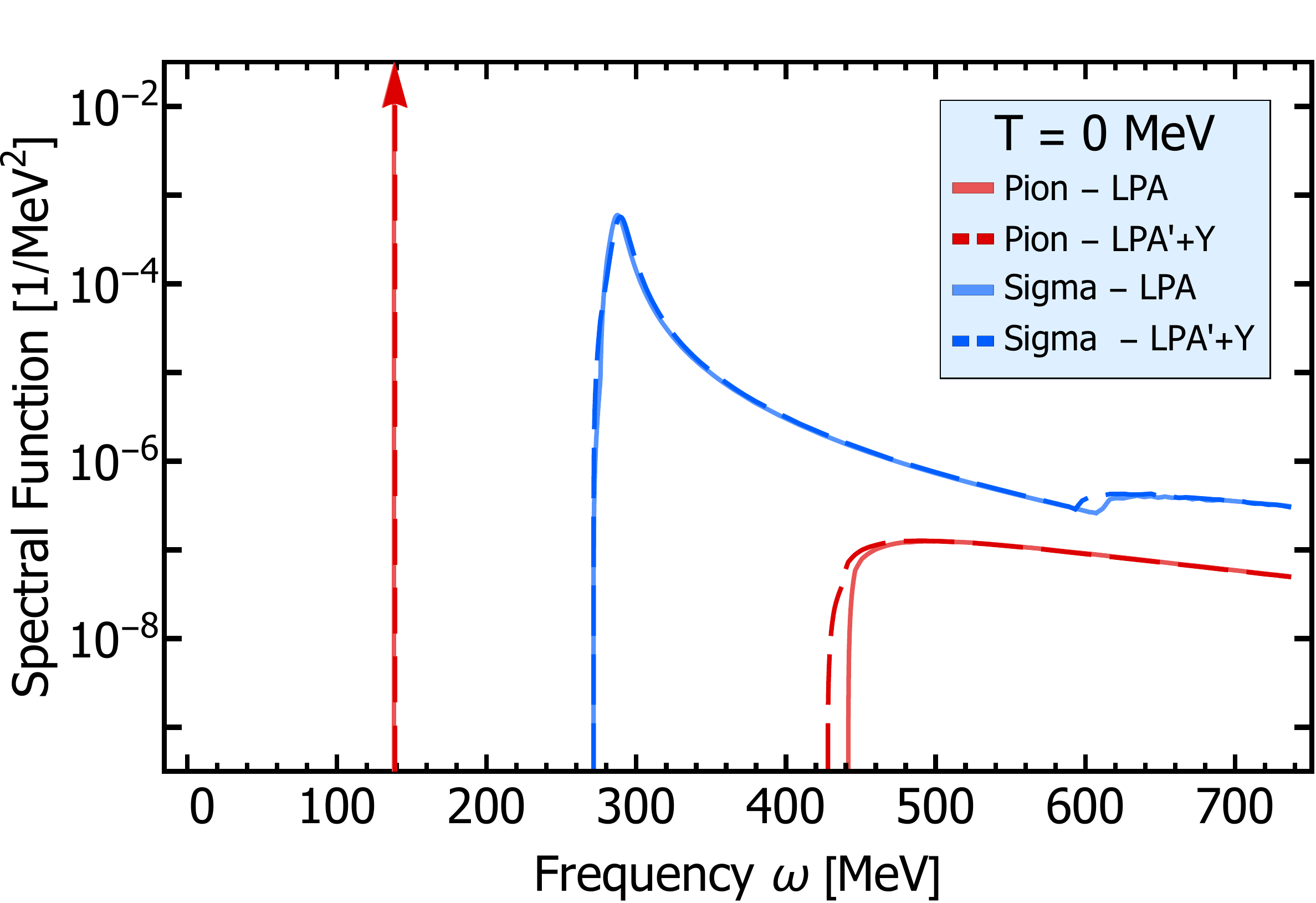} &   
		\includegraphics[width=0.5\linewidth]{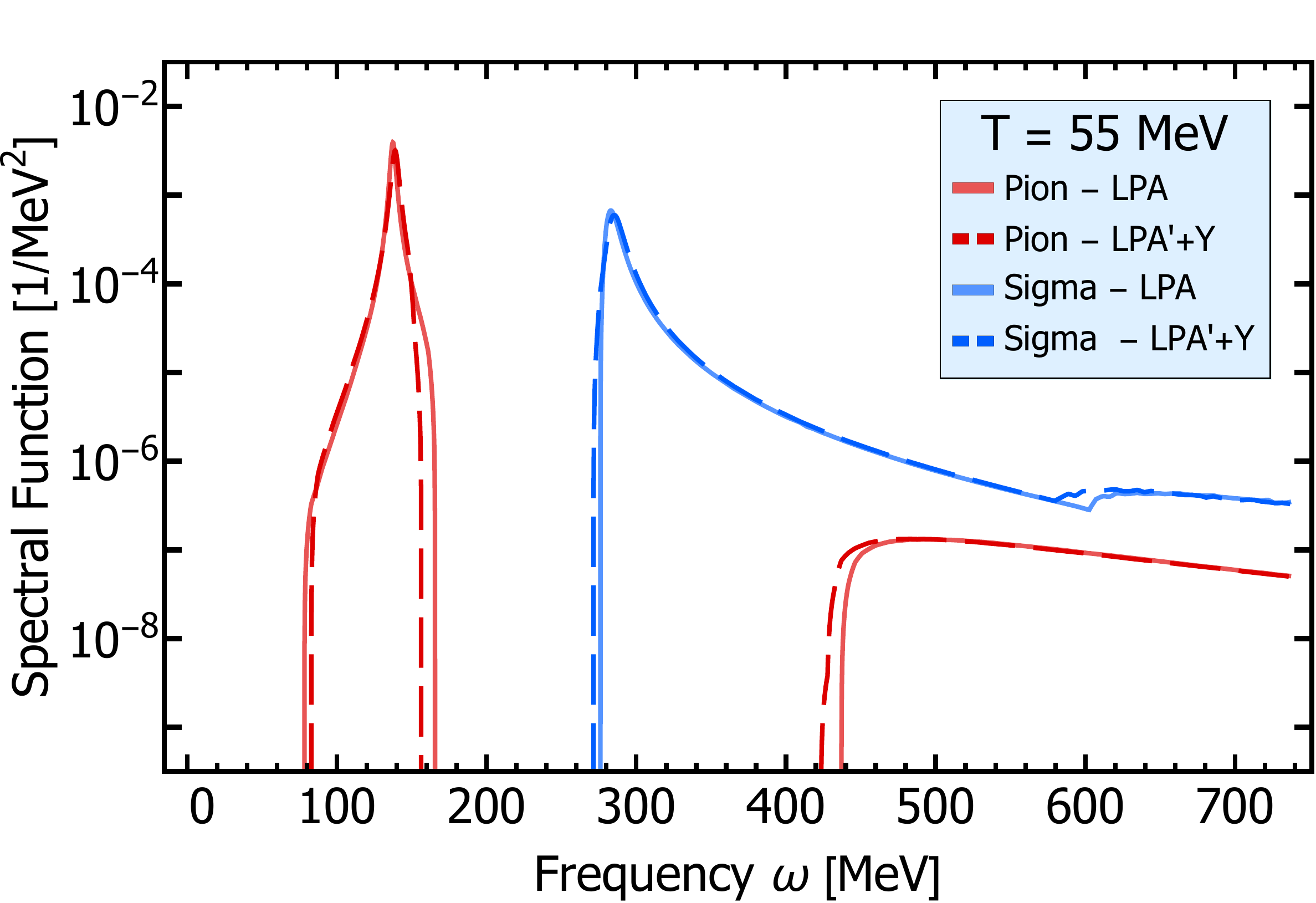}\\ T=0 MeV\  & T=55 MeV\\
		\includegraphics[width=0.5\linewidth]{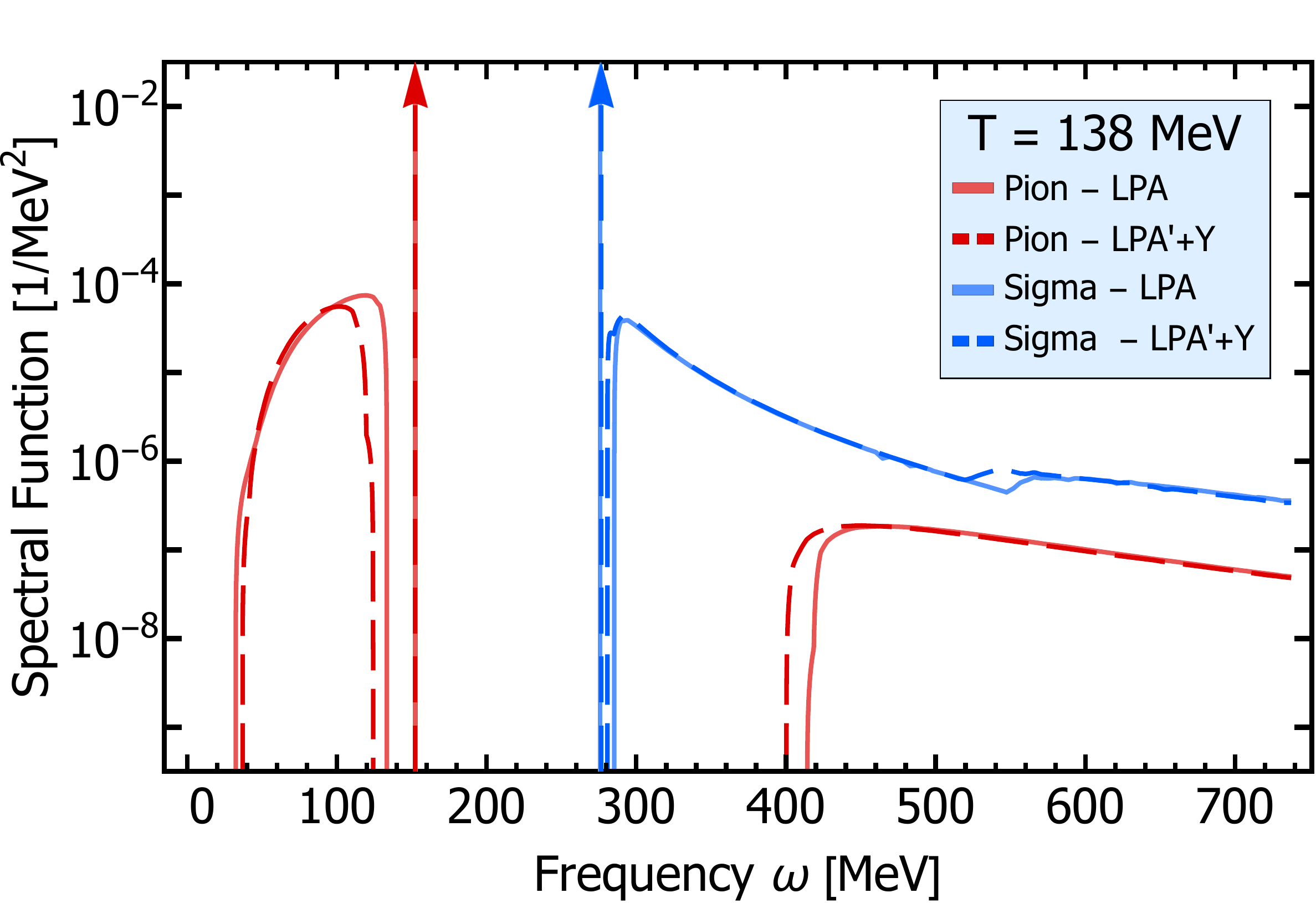} &   
		\includegraphics[width=0.5\linewidth]{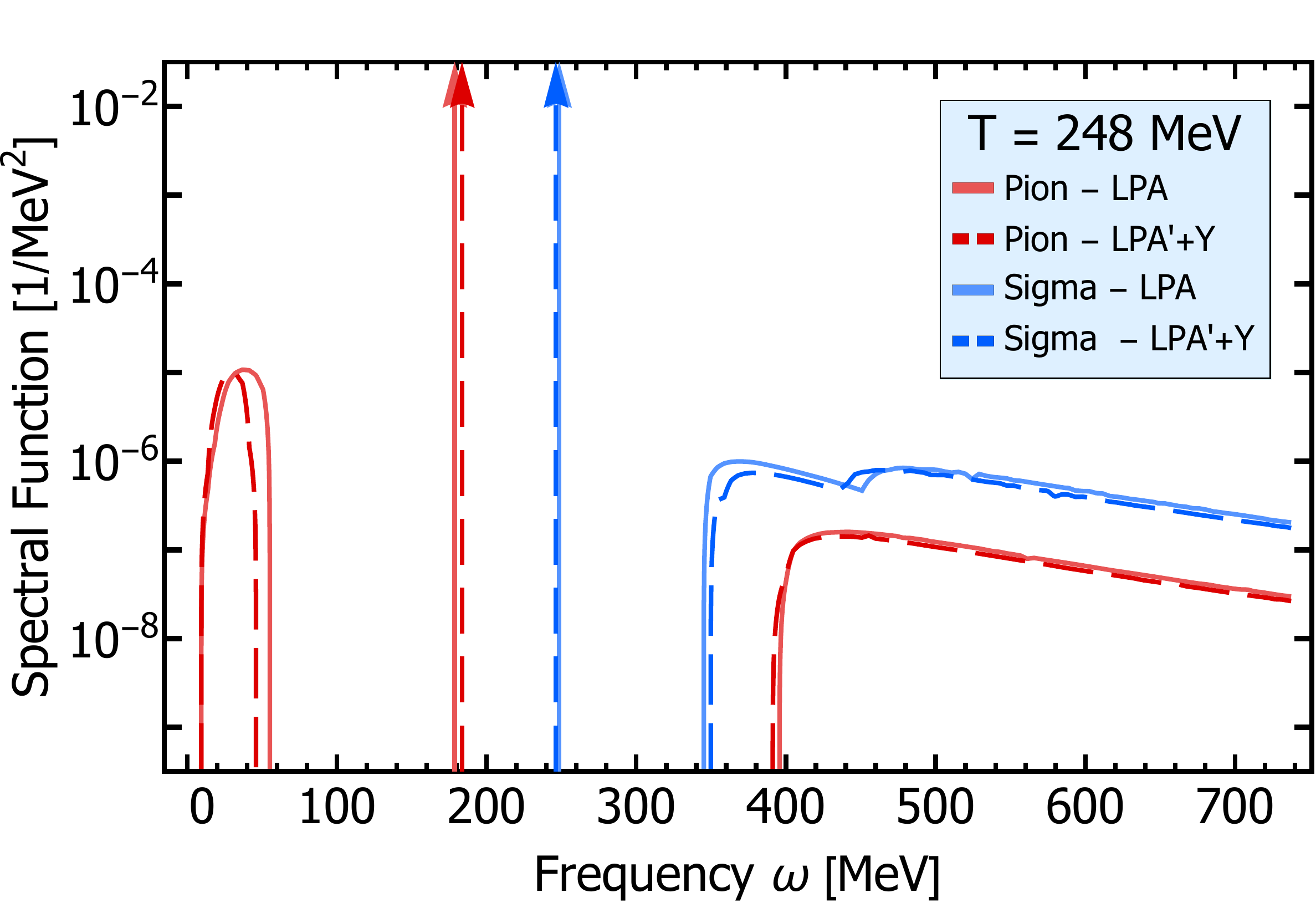}\\ T=138 MeV\  & T=248 MeV\\
	\end{tabular}
	\caption{Spectral functions at different temperatures comparing LPA to LPA'+Y.}
	\label{fig:LPApYspec}
\end{figure*}
\begin{figure*}
	\begin{tabular}{c c}
		\includegraphics[width=0.5\linewidth]{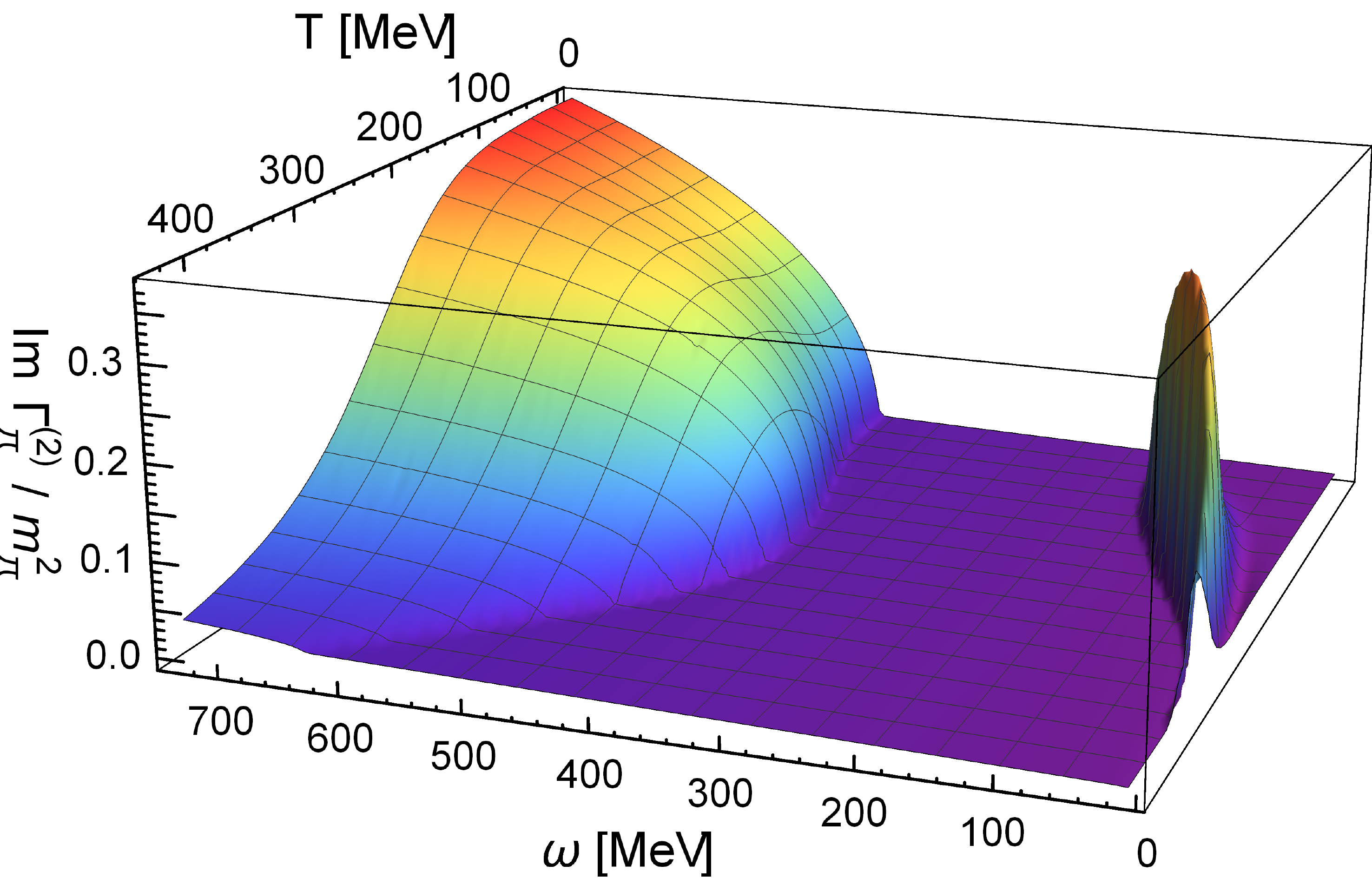} &
		\includegraphics[width=0.5\linewidth]{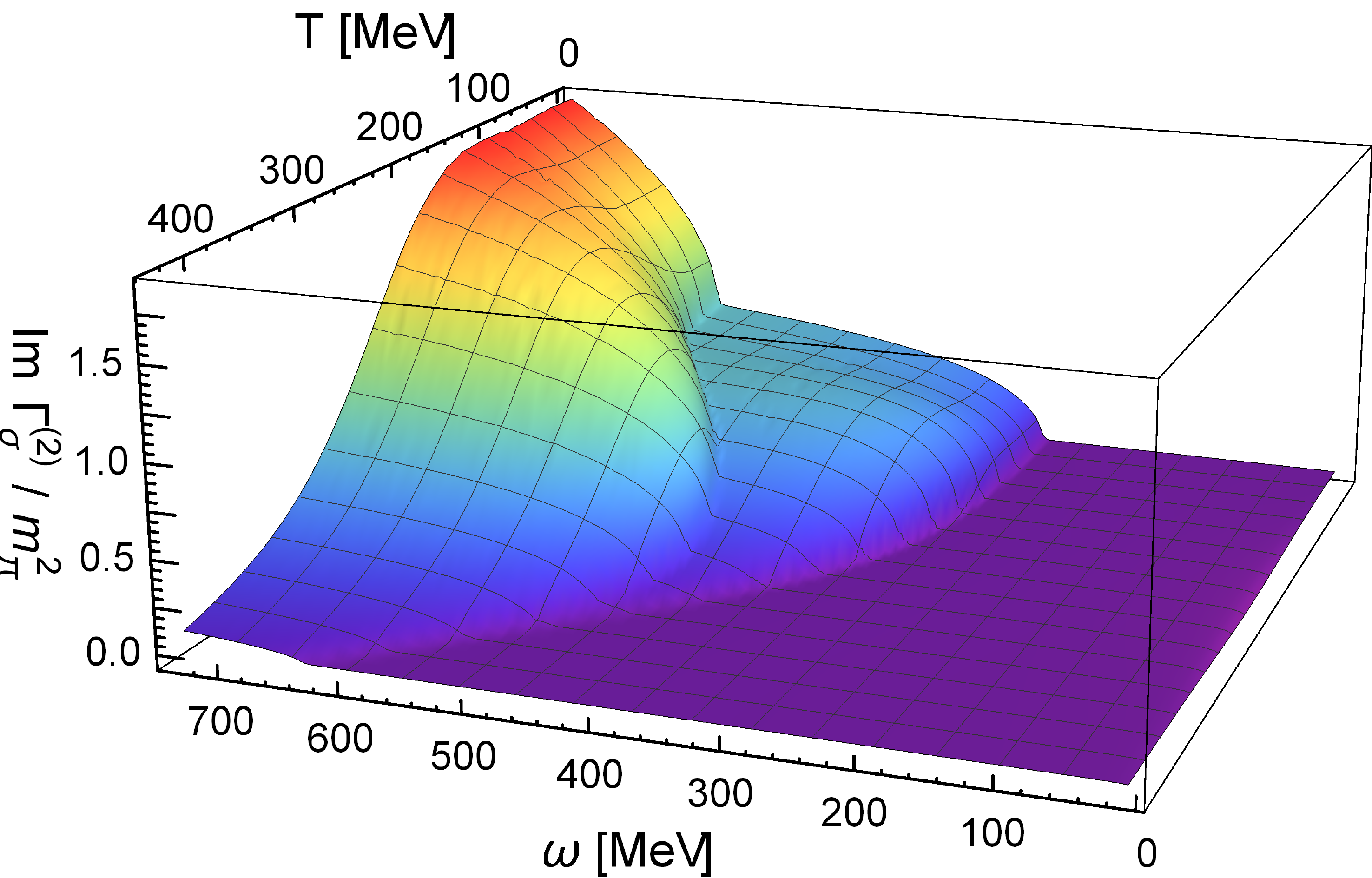}
	\end{tabular}
	\caption{Cut structure of the propagators in the frequency-temperature plane.}
	\label{fig:imparts}
\end{figure*}
Turning to finite temperature, it turns out that the simple evaluation
of the Matsubara sum in the analytically continued expression does
again not lead to the desired contribution to the self-energy of the
retarded propagator. As demonstrated in \cite{Pawlowski:2015mia} for a
propagator with only simple poles, the difference between the two
results can be written as a sum of residues corresponding to all poles
of the propagator involving the external frequency weighted by a
difference of thermal occupation numbers.  This generalizes
straightforwardly to the situation with propagators with a general
complex structure: The correction term is obtained by evaluating
contour integrals surrounding all complex structures of the propagator
involving the external frequency weighted with a difference of
occupation numbers $n(\imag(p_0+ q_0))-n(\imag q_0)$.  This is
illustrated in Fig.~\ref{fig:continuationT}. This procedure represents
a formal solution of the analytical continuation problem in the
propagator approximation. The numerical implementation of this
procedure turns out to be very challenging as it requires to evaluate
contour integrals in the close vicinity of first- and second-order
poles (as soon as the regulator term $G\dot R G(q)$ is introduced). In
this work we restrict ourselves to the evaluation of finite
temperature spectral functions on a given LPA'+Y solution. In this
case the propagators show only simple poles and the correction term
can be evaluated as a sum of residue contributions instead of an
actual contour integration, further details can be found
in~\App{a:numerics}.

Technically, the retarded two-point correlator is obtained from the
Euclidean correlation function:
\begin{align}\label{eq:q:cont_correlator}
  \Gamma^{(2)}_R(\omega,\vec{p}) = \lim_{\varepsilon \to 0} 
  \Gamma^{(2)}_\text{\tiny{Eucl}}(p_0=-\imag (\omega + \imag \varepsilon), \vec{p})\, ,
\end{align}
which in turn can then be used to obtain the spectral function
\begin{align} \label{eq:spectral_function} \rho(\omega,\vec{p}) = 2\
  \text{Im}\, G_R(\omega,\vec{p}) Z(0)\, ,
\end{align}
where $Z(0)$ is the wave function renormalization at vanishing
momentum. The spectral function in \eq{eq:spectral_function} is
renormalization group invariant, but is not normalised to one,
\begin{align}\label{eq:rhonorm}
          \int_{\mathbb{R}^+} d \omega^2\,  \rho(\omega,0) = \mathcal{N}_\rho\,.
\end{align} 
This comes from the normalisation with $Z(0)$. Typically, the physical
spectral function is normalised to one,
\begin{align}\label{eq:rhophys}
	\hat \rho = \frac{1}{\mathcal{N}_\rho }\rho\,.
\end{align}	
	
%
\subsection{Truncations}
In order to meet the requirement of only having to take pole
corrections into account when calculating spectral functions, we
restrict our effective action to the following form
\begin{equation}\label{eq:trunc}
  \Gamma_k = \int \mathrm{d}^4 x\ \left( \frac{Z}{2}\left(\partial_\mu \phi_a
    \right)^2 + \frac{Y}{8}\left(\partial_\mu \rho\right)^2 + V(\sigma) \right) \, ,
\end{equation}
with the mesonic field $\phi=(\sigma,\vec{\pi})$.  The effective
potential, $V(\sigma)$, is split into a part only depending on the
$O(4)$-invariant $\rho=\frac{1}{2}(\pi_a^2+\sigma^2)$, and a part
explicitly breaking the symmetry, thus allowing for the Goldstone
bosons to acquire a finite mass
\begin{equation}
	V(\sigma) = U(\rho) - c \sigma \, .
\end{equation}
Note that the linear breaking term drops out of all dynamical
equations, and in particular the flow equations. While this fact is
hidden in an expansion about the flowing minimum, it is apparent
within an expansion around a fixed expansion point in the bare field
$\sigma$, see \cite{Pawlowski:2014zaa}. In the present work we choose
the latter expansion for both stability and for a self-consistent
frequency dependence, see \cite{Helmboldt:2014iya}.
We choose the expansion point close but above the
IR minimum $<\!\!\sigma \!\!>$, as argued in
~\cite{Pawlowski:2014zaa}. We take into account terms up to
$\phi^{14}$, and the convergence of the results has been checked.

The truncation~\eq{eq:trunc} is refered to as LPA'+Y and we
additionally consider a second truncation with $Z=1$, $Y=0$
corresponding to the usual LPA (Local Potential Approximation) scheme.
Effectively the two dressing functions $Z$ and $Y$ are reabsorbed into
a dressing for the pion, $Z_\pi$, and one for the sigma meson,
$Z_\sigma$.
	
Finite temperature is introduced via the Matsubara formalism in the
usual manner.
	
As argued in~\cite{Pawlowski:2015mia}, we are mostly concerned with
regulators depending on the Lorentz invariant momentum configurations
$q^2$ and is discussed at large in said reference. Further details
about the regulator in the current work are given in~\App{a:reg} and a
comparison with a Lorentz invariance breaking regulator is discussed
in~\App{a:RegCompare}.
\section{Results}
We first display the temperature and momentum dependence of the
spectral functions. Furthermore, we extract the pole mass and compare
it to the curvature mass. Details about the numerical implementation
are given in~\App{a:numerics}.
\subsection{Spectral functions at vanishing external
  momentum} \label{sec:spectral_result}
Here we show the temperature
evolution of the pion and sigma spectral function in the LPA'+Y
approximation.
The spectral functions feature several district structures with a
clear physical interpretation, see e.g.~\cite{Tripolt:2014wra} for a
detailed discussion of the different processes in the Quark--Meson
model. In general there are two different cut structures at finite
temperature and vanishing external momentum in the propagator extended
to the complex plane.
The unitarity cut spanning from the multi-particle decay threshold to
infinity, i.e. $\omega\in[\mu_\text{thresh},\infty)$, which is present
in any interacting theory
Furthermore, the Landau cut at smaller
frequencies, which is purely medium dependent
and gives rise to inverse scattering processes \cite{Weldon:1983jn}
with the heat bath. These scattering processes give
rise to Landau damping, hence the name.
Finally, delta functions represent stable
particles.
\begin{figure*}[t]
	\centering
	\begin{subfigure}[b]{.47\linewidth}
		\includegraphics[width=\linewidth]{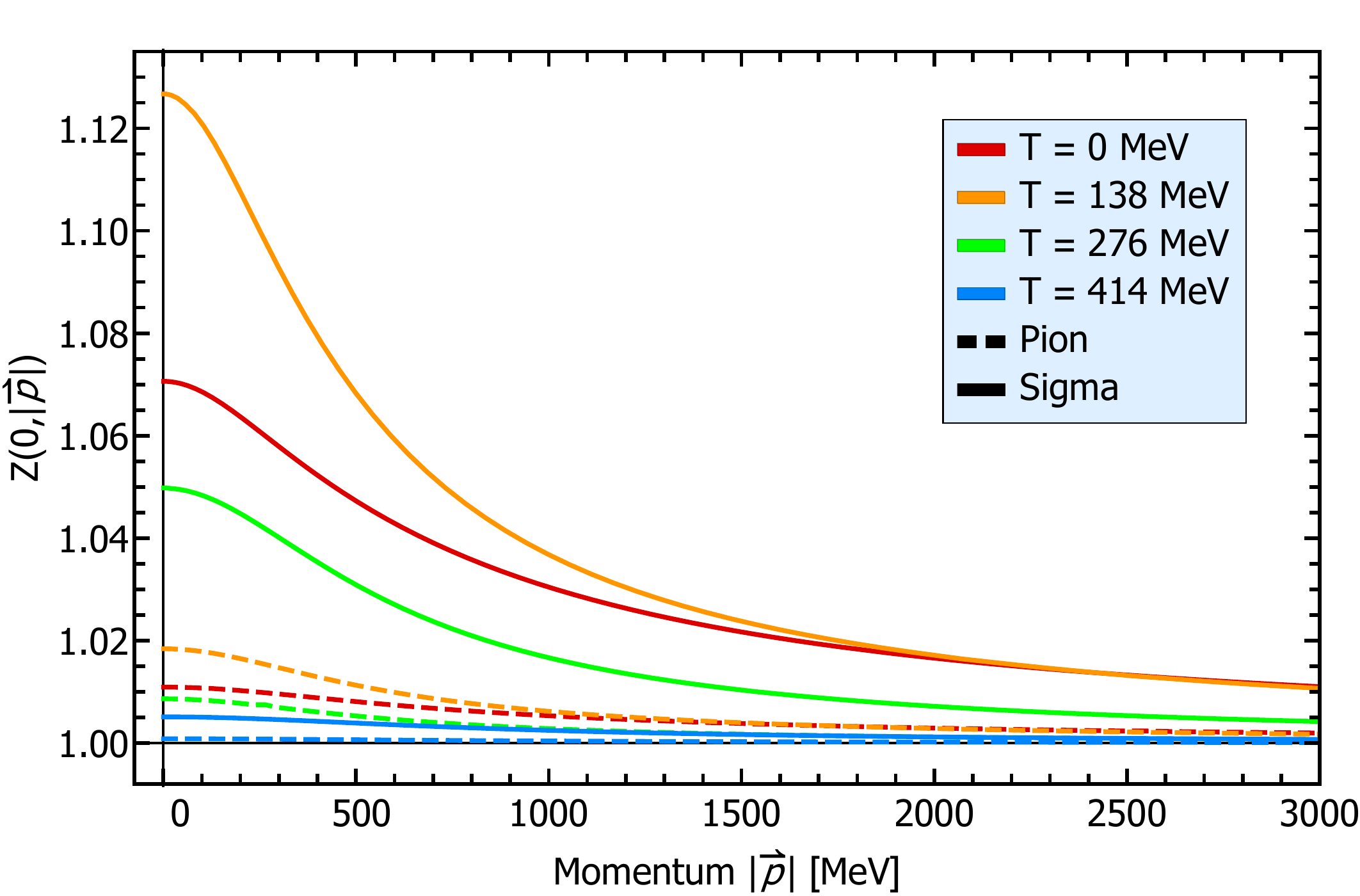}
		\caption{Momentum dependent Euclidean dressings.\\ \phantom{AAAAAAAA}}\label{fig:euclMomDep}
	\end{subfigure}
	\begin{subfigure}[b]{.45\linewidth}
		\includegraphics[width=\linewidth]{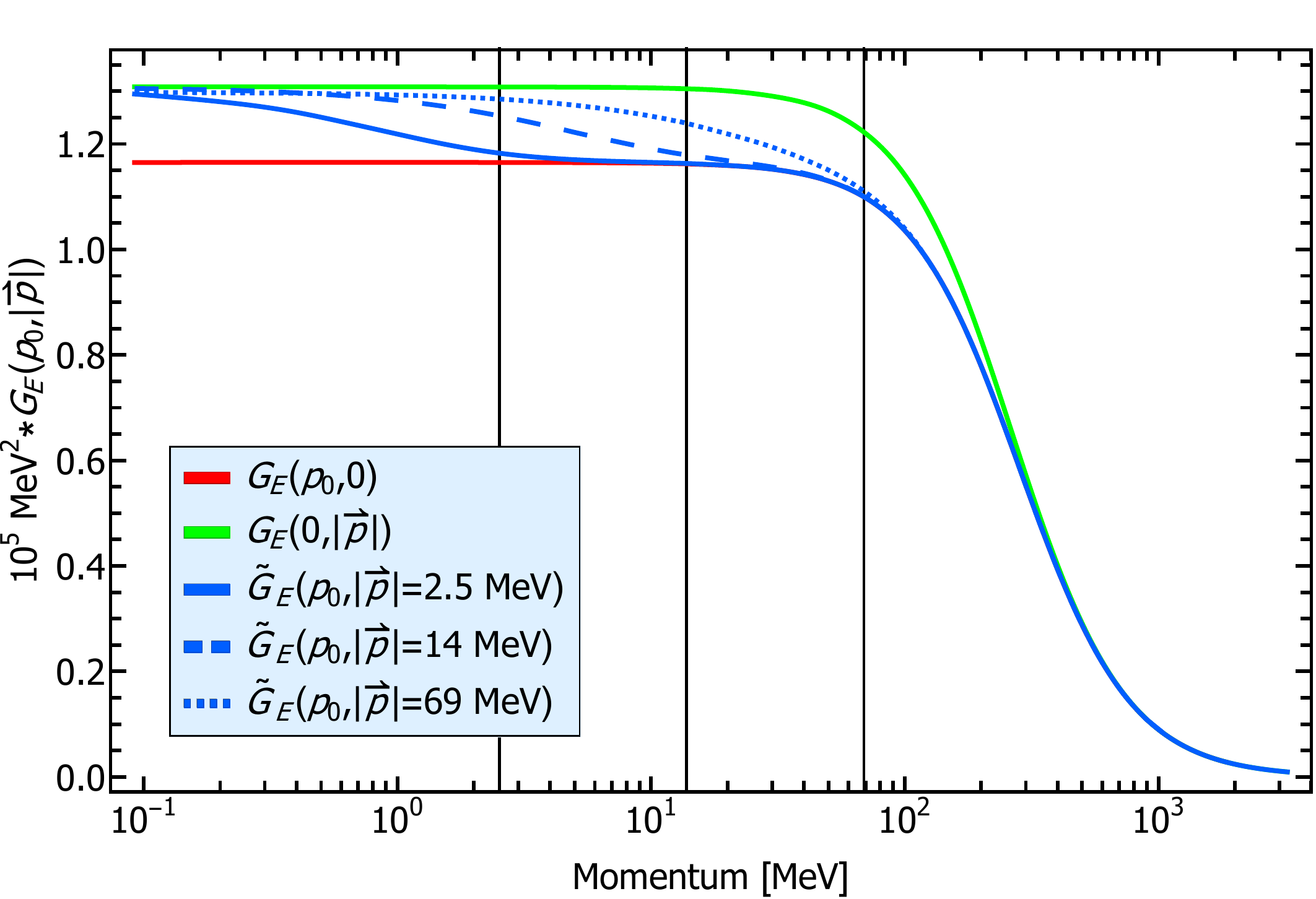}
		\caption{Non commuting limits~\eq{eq:diff_limits} shown for the sigma propagator at  $T = \unit[138]{MeV}$.}\label{fig:diffLimitsPlot}
	\end{subfigure}
	
	\begin{subfigure}[b]{.45\linewidth}
		\includegraphics[width=\linewidth]{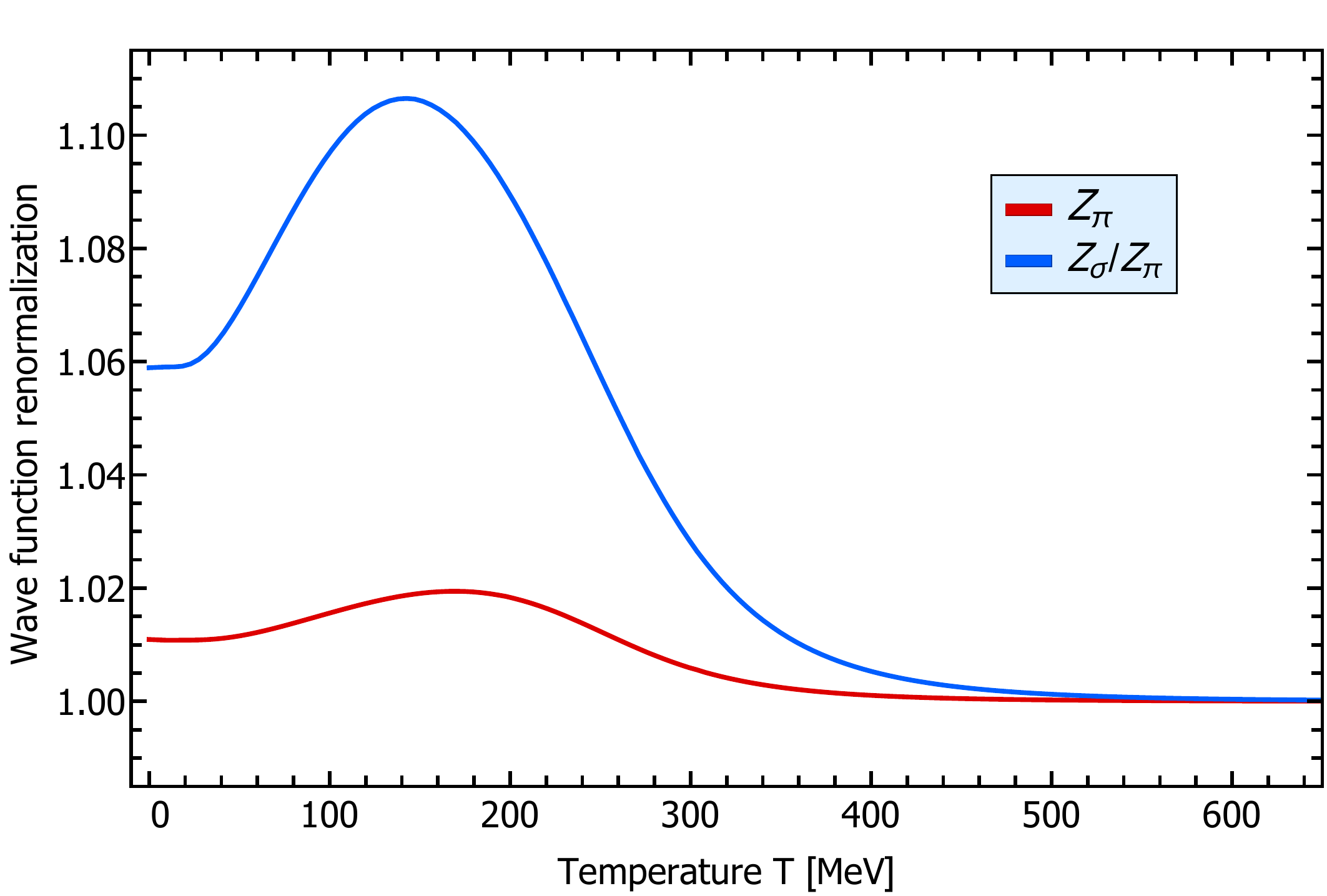}
		\caption{Wave function renormalizations.}\label{fig:ZsigmaPlot}
	\end{subfigure}
	\begin{subfigure}[b]{.45\linewidth}
		\includegraphics[width=\linewidth]{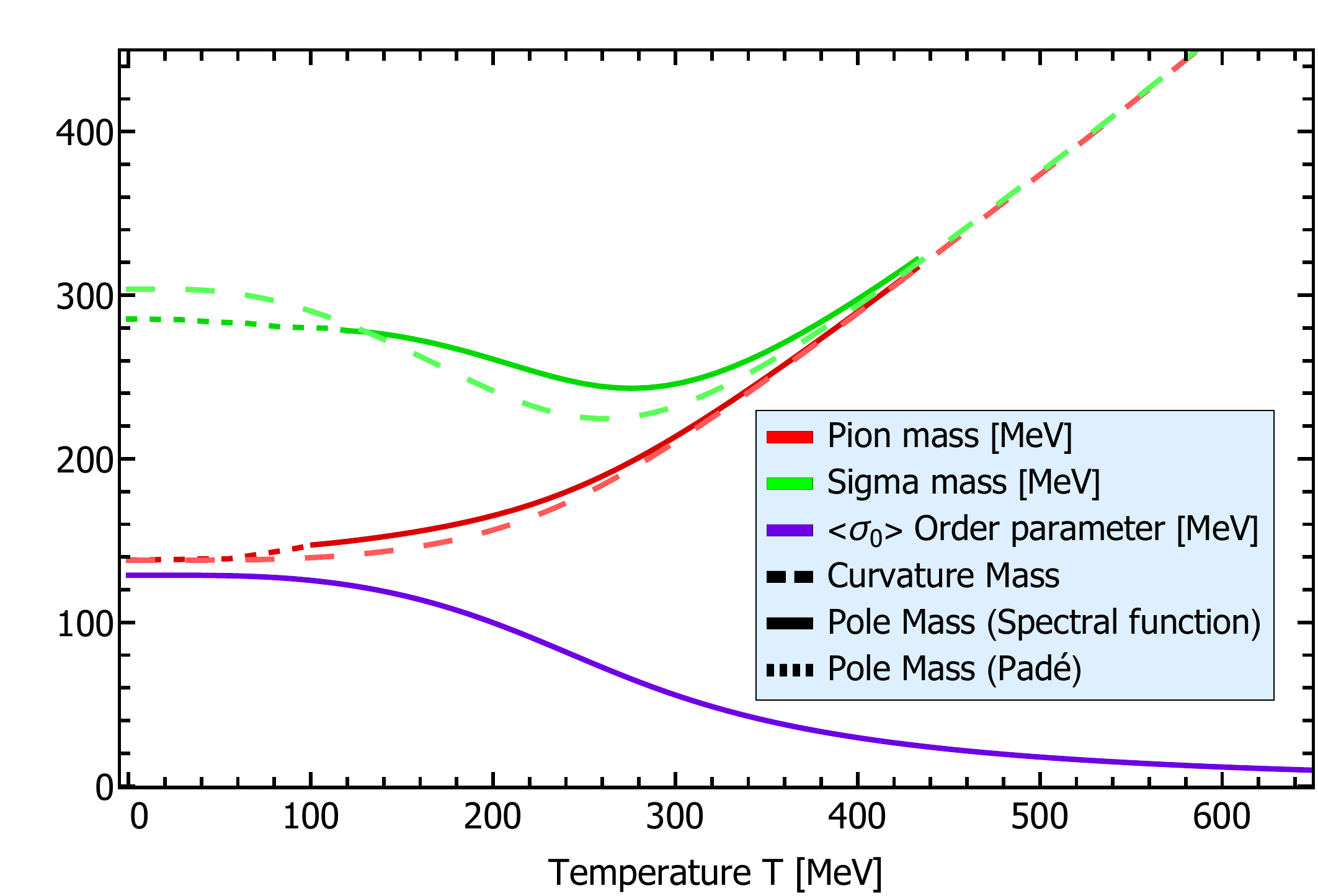}
		\caption{Different masses and order parameter.}\label{fig:mass_plot}
	\end{subfigure}
	\caption{Aspects concerning the relation between pole and curvature mass.}
	\label{fig:MassRelated}
\end{figure*}
\begin{figure}[t]
	\includegraphics[width=\linewidth]{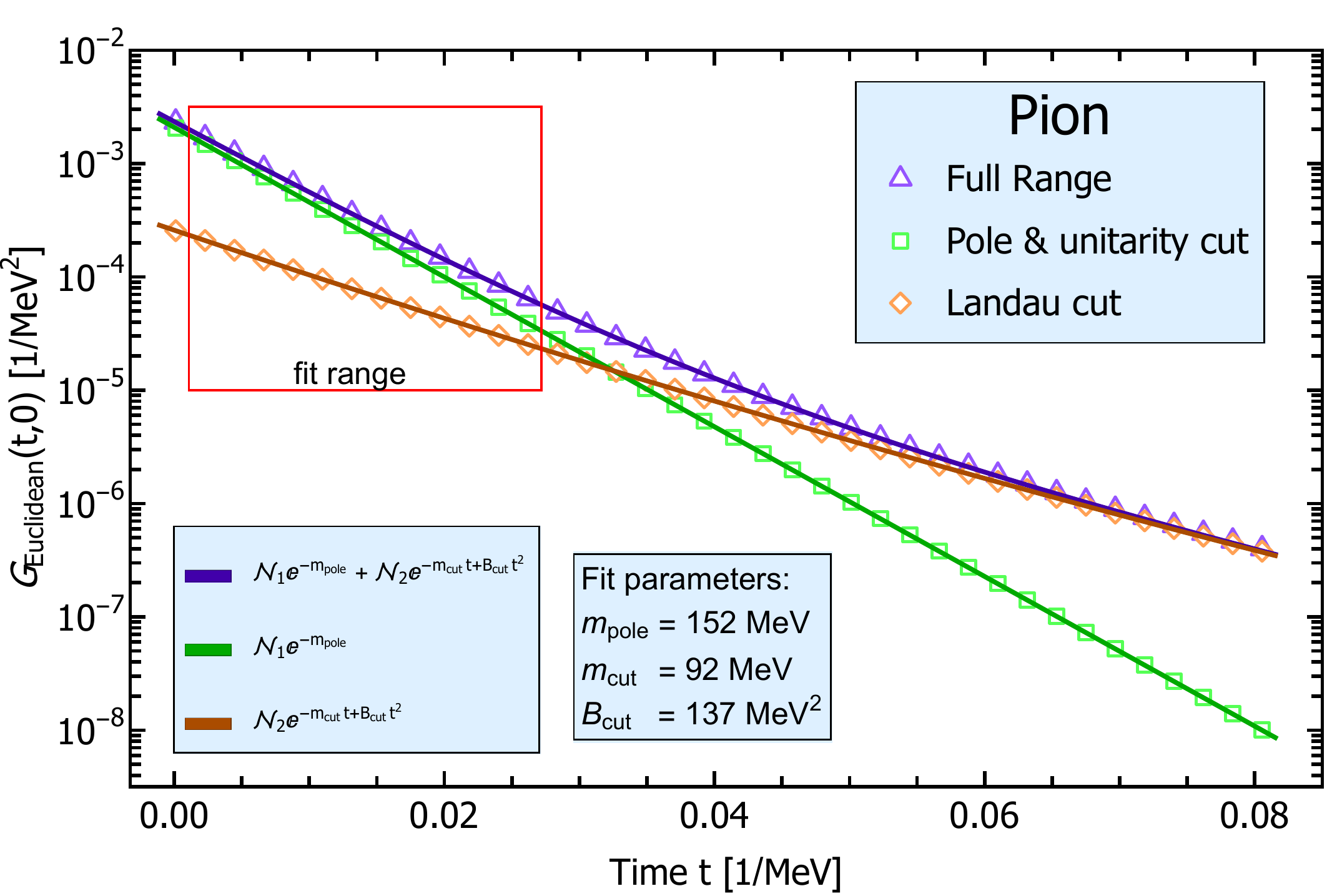}
	\caption{Fourier transformed pion propagator.}
	\label{fig:FourierPlot}
\end{figure}
Our result for the spectral functions at vanishing external momentum
in the LPA and LPA'+Y truncation are depicted
in~\Fig{fig:LPApYspec}. For the sigma meson there are two different
processes available, i.e. $\sigma^*\rightarrow \pi+\pi$ and
$\sigma^*\rightarrow \sigma+\sigma$ and no Landau cut structure. While
for the pion we have $\pi^*\rightarrow\pi+\sigma$ for the Unitarity
cut and $\pi^* + \pi \rightarrow \sigma$ for the Landau cut. At
vanishing temperature we have the expected stable pion and no stable
sigma particle. At finite temperature the sigma meson emerges as
stable particle as O(4)-symmetry gets restored. The presence of stable
particles at finite temperatures is an artefact of the current
truncation.

The difference between the two truncation is most prominently seen at
multiple particle decay thresholds, which involve a sigma meson
demonstrating the effect of the wave-function renormalization, shown
in~\Fig{fig:ZsigmaPlot}, on the curvature mass, as in both truncations
the proper pole mass is not coupled back into the system. A problem
that is numerically not traceable in the current formalism,
c.f. discussion in~\App{a:numerics}.  The cut structure of propagators
is best seen in the imaginary part of the two-point function,
$\text{Im } \Gamma^{(2)}(\omega)$, as a finite value translates
directly into a cut of the propagator, the result is shown
in~\Fig{fig:imparts}.

The finite part at small frequencies, that vanishes for
larger temperatures, in the pion shows again the Landau cut. For larger
frequencies the unitarity cut shows clear decay thresholds, i.e. in
the sigma meson the different thresholds and their degeneracy for high
temperatures can be seen nicely.

Our results compare qualitatively well to the results obtained using a spatially flat regulator, c.f. the discussion in \App{a:RegCompare}, in
the Quark-Meson model \cite{Tripolt:2013jra, Yokota:2016tip, Wang:2017vis, Yokota:2017uzu}.
\subsection{Pole, screening \& curvature masses} \label{sec:masses}
It is interesting to compare the real time pole and screening masses
with the Euclidean curvature mass, for a detailed discussion of the
respective definitions in the present FRG context see
e.g.~\cite{Helmboldt:2014iya}.  The curvature mass is typically used
in Euclidean computations within low energy effective field theories
for QCD. There, the physical pion and sigma masses are input
parameters and are identified with the respective curvature
masses. However, they have to be identified with the pole mass, and
hence we have to check how well such an identification works. 
\begin{figure*}
	\begin{tabular}{cc}
		Pion\  & Sigma\\
		\includegraphics[width=0.5\linewidth]{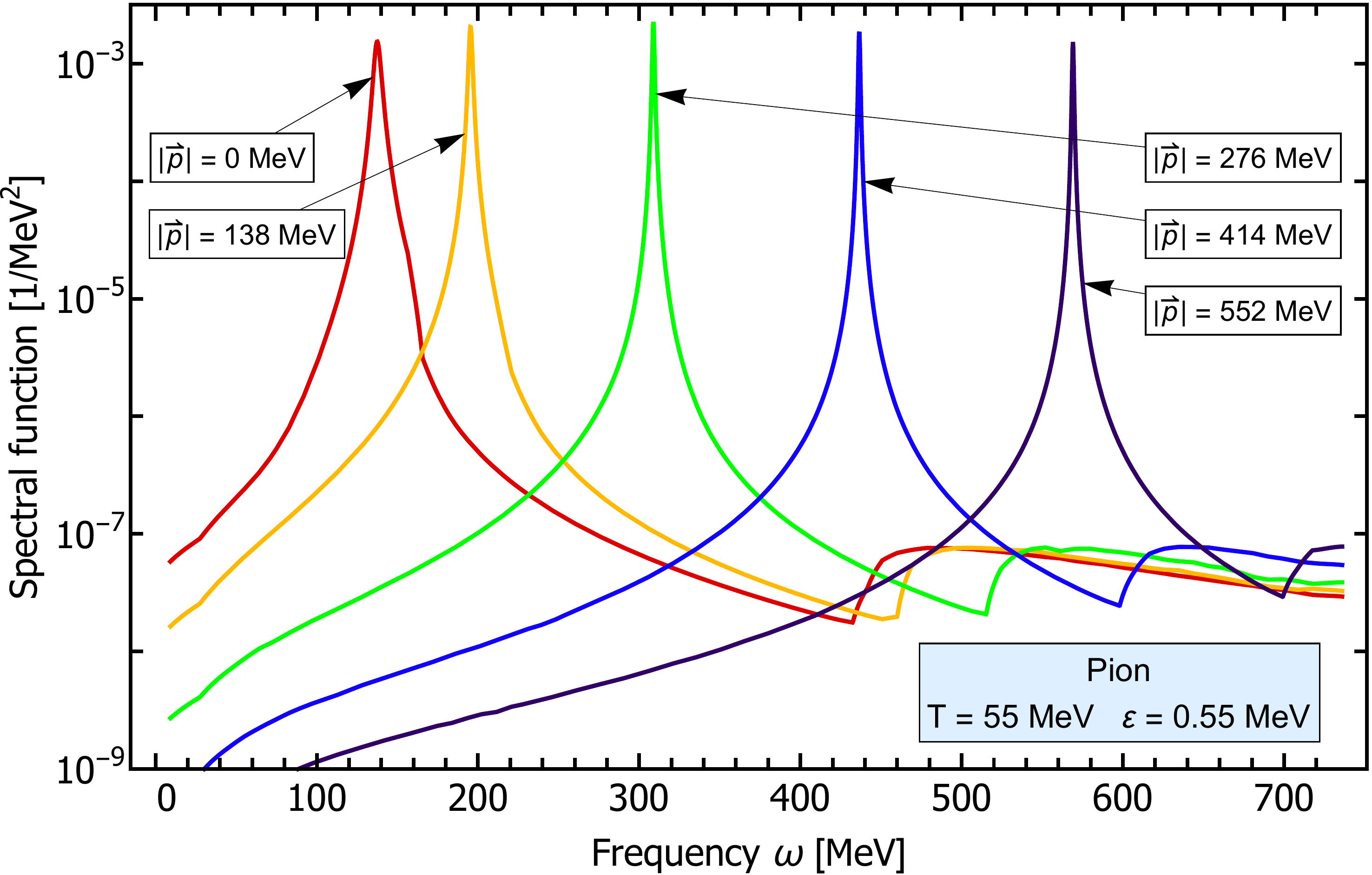} &   \includegraphics[width=0.5\linewidth]{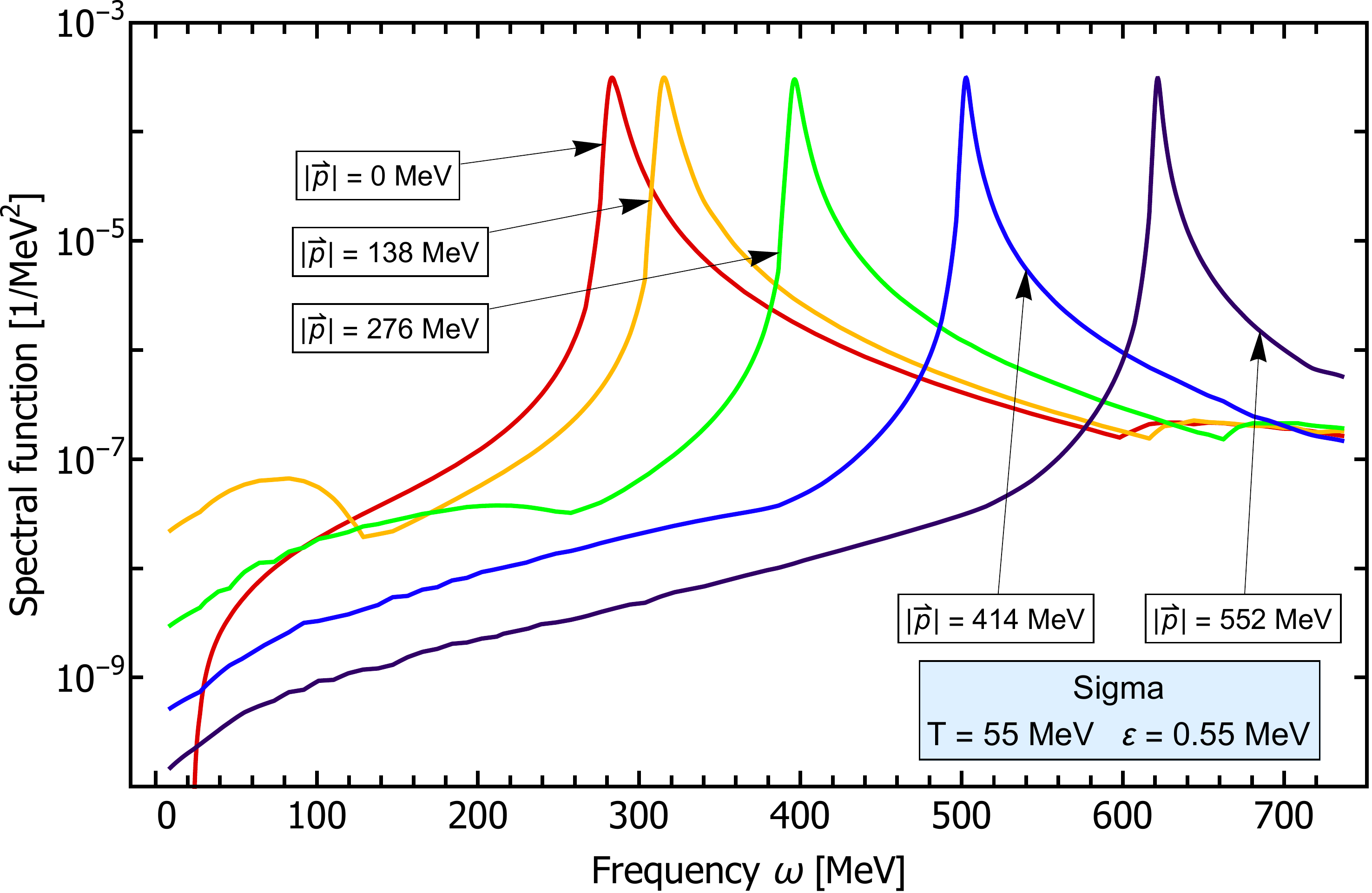}\\ 
		\includegraphics[width=0.5\linewidth]{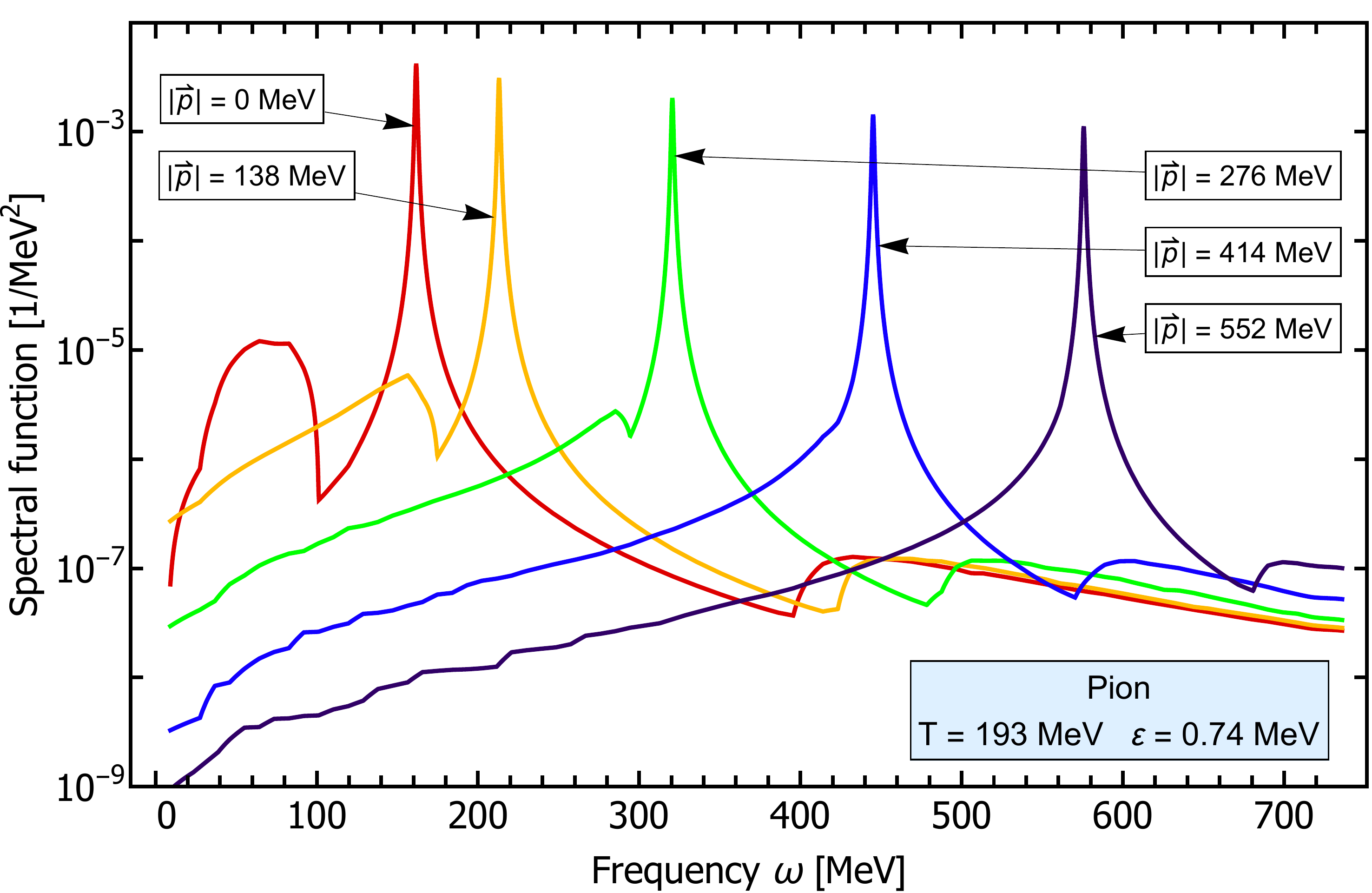} &   \includegraphics[width=0.5\linewidth]{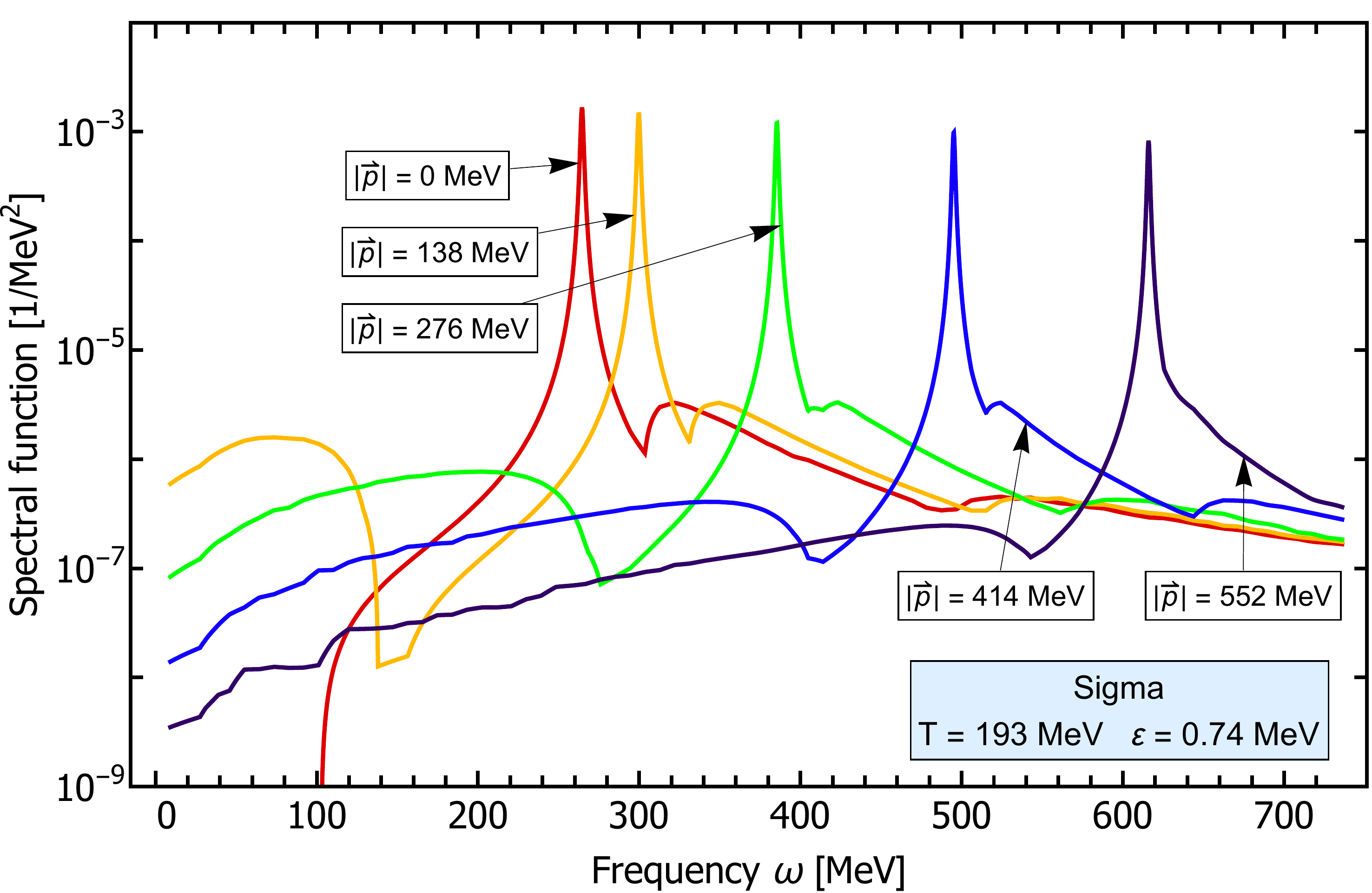}\\ 
	\end{tabular}
	\caption{Spectral functions at finite external momenta for different temperatures across the phase transition.}
	\label{fig:ExtMom}
\end{figure*}
In the case of a stable particle, the position of the pole can be
directly extracted from the spectral function. For particles with a
finite decay width, this is not possible any more as the pole leaves
the physical sheet of the complex energy plane and is found on the
second Riemann sheet.

On the other hand the pole mass $m_{\text{\tiny{pole}}}=1/\xi_t$ is the
inverse temporal screening length, which can be extracted from
\begin{align}\label{eq:pole}
\lim_{t\to \infty}  G_E(t, 0)\propto  e^{-t/\xi_t} \,.
\end{align}
Furthermore the screening mass is
$m_{\text{\tiny{screen}}}=1/\xi_{\text{\tiny{spat}}}$, that is the
inverse spatial screening length,
\begin{align}\label{eq:screening}
\lim_{|\vec x|\to \infty}  G_E(0, \vec x)\propto  e^{-|\vec x|/\xi_{\text{\tiny{spatial}}}} \,,
\end{align}
and the curvature mass is 
\begin{align}\label{eq:cur}
m^2_{\text{\tiny{cur}}}= \lim_{\vec{p}\to 0} \frac{\Gamma^{(2)}_E(0,\vec{p})}{Z(0,\vec{p})}\,. 
\end{align}
Note that the wave function renormalisation $Z$ in \eq{eq:cur} ensures
the RG-invariance of the latter. However, the necessity of a choice of
the momentum configuration (in \eq{eq:cur} it is $p_0=0,\vec p\to 0$)
introduces a scheme dependence, while the pole and screening masses
are scheme-independent.

Note also that at finite temperature the order of limits in~\eq{eq:cur} matters, as
the two limits do not commute any more
\begin{align} \label{eq:diff_limits}
\lim_{\vec{p}\to 0} \lim_{p_0\to 0} \Gamma^{(2)}(p_0,
\vec{p}) \neq \lim_{p_0\to 0}\lim_{\vec{p}\to 0} \Gamma^{(2)}(p_0,\vec{p})\, ,
\end{align}
where the order of limits on the LHS is referred to as \textit{static}
and the limit on the RHS as \textit{plasmon}, for details see
e.g.~\cite{das1997finite}, if not stated otherwise $Z(0,0)$ is
understood in the \textit{static} limit. The difference between the
two limits can be seen easily in the corresponding spectral functions,
as the static limit contains the transport peak, while this structure
is absent the \textit{plasmon} limit, compare e.g.~\Fig{fig:LPApYspec}
with~\Fig{fig:ExtMom} and the see the corresponding discussion
in~\Sec{sec:finite_ext_mom}.
\begin{figure*}
	\begin{tabular}{cc}
		\includegraphics[width=0.5\linewidth]{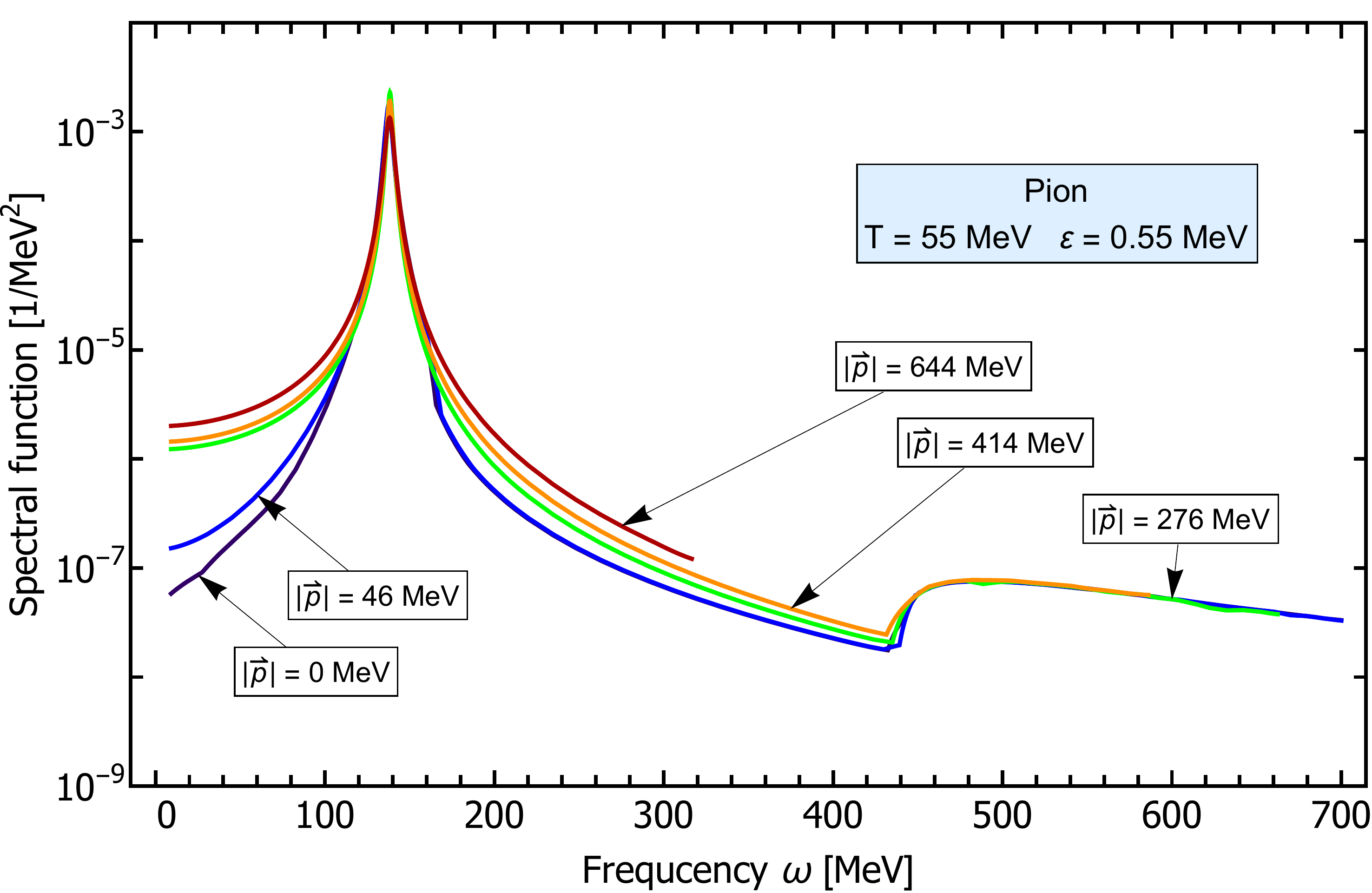} &   \includegraphics[width=0.5\linewidth]{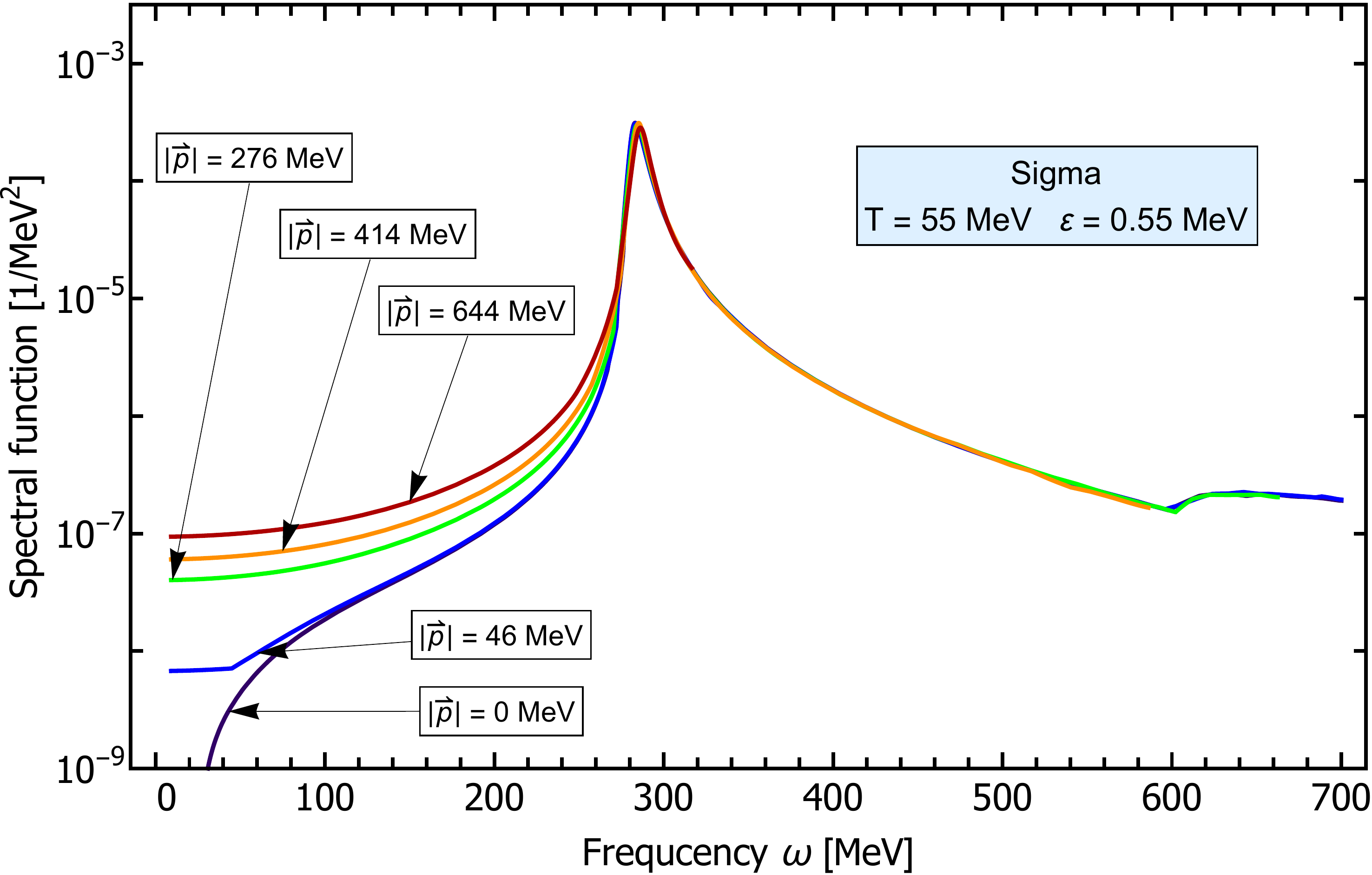}\\
	\end{tabular}
	\caption{Lorentz invariance of the spectral functions, $\rho(\sqrt{\omega^2+\vec{p}^{\, 2}},\vec{p})$, at a small temperature.}
	\label{fig:lorentzInvariance}
\end{figure*}
The difference between the two limits
can also be seen in the corresponding Euclidean correlation functions, shown in~\Fig{fig:diffLimitsPlot},
where the green line shows the propagator as a function of the external momentum at the zeroth Matsubara mode,
i.e. approaching \textit{static} limit. The red line shows the propagator as a function of the external Euclidean frequency,
continuously defined due to the analytic continuation involved, which approaches the \textit{plasmon} limit. As soon as a
small external momentum is introduced, the \textit{static} limit is implicitly taken and the scale is given by the external momentum, shown as vertical lines. For better visibility the trivial mismatch for larger $\vec{p}$ has been subtracted, $\tilde{\Gamma}^{(2)}(p_0,\vec{p}\,) = \Gamma^{(2)}(p_0,\vec{p}\,) - \vec{p}^2$ and $\tilde{G}=\tilde{\Gamma}^{-1}$.

We can now use~\eq{eq:cur} together
with~\eq{eq:spectral_euclidean} together with the spectral
representation of the Euclidean propagator
\begin{align}\label{eq:spectral_euclidean}
Z(0,0) G_E(p_0,\vec{p})=\int  \mathrm{d}\eta\ \frac{\rho(\eta, \vec{p})}{\eta+\imag p_0}
\end{align}
to derive a relation between the curvature mass and the spectral representation
\begin{align}
m^2_{\text{\tiny{cur}}} &= \frac{\int \frac{\mathrm{d}\eta}{\eta} \hat\rho(\eta,0)
}{
	\int \frac{\mathrm{d}\eta}{\eta}\frac{1}{\eta^2} \hat\rho(\eta,0)}\,, 
\end{align} 
with the normalised spectral function \eq{eq:rhophys}.  In case a
stable particle is present, the spectral function can be split in two
positive parts as follows,
\begin{align}\label{eq:rhosplit}
&\hat\rho = \hat\rho_{\text{\tiny{pole}}} + \hat \rho_{\text{\tiny{cut}}}\,,  \nonumber \\
&\hat \rho_{\text{\tiny{pole}}} = \sign(\omega) Z_{\text{\tiny{pole}}} \delta(p^2-m^2_{\text{\tiny{pole}}})\,.
\end{align}
The normalisation of the pole is smaller than one:
$Z_{\text{\tiny{pole} }}= 1- \int d\omega^2
\hat\rho_{\text{\tiny{cut}}}<1$ which follows from the normalisation
of the spectral function and the positivity of
$\rho_{\text{\tiny{cut}}}$. Using this split in
\eq{eq:spectral_euclidean} we arrive at
\begin{align}\label{eq:polcur}
m^2_{\text{\tiny{cur}}} &=m_{\text{\tiny{pole}}}^2\frac{Z_{\text{\tiny{pole}}}  + m_{\text{\tiny{pole}}}^2  
	\int \frac{\mathrm{d}\eta}{\eta} \rho_\text{\tiny{cut}}(\eta,0)
}{
	Z_{\text{\tiny{pole}}}  + m_{\text{\tiny{pole}}}^4  \int \frac{\mathrm{d}\eta}{\eta} \frac{1}{
		\eta^2} \rho_\text{\tiny{cut}}(\eta,0)
}\, ,
\end{align}
\Eq{eq:polcur} entails the information when the difference between
pole and curvature masses grows large: Firstly, decreasing
$Z_{\text{\tiny{pole}}}$ increases the importance of the cut-part and
hence the difference between curvature and pole mass grows. Secondly,
if the spectral weight of $\rho_{\text{\tiny{cut}}}$ is taken at
smaller spectral values, the integrals in \eq{eq:polcur} grow and
hence the difference between curvature and pole mass grows.  

Translated back to Euclidean space-time, both processes lead to strong
frequency and momentum dependences in the Euclidean propagator. In
turn, if the pole term dominates the full spectral function, the full Euclidean propagator 
is well described by a propagator with a constant wave function renormalization, depicted in~\Fig{fig:ZsigmaPlot}.  
A similar conclusion was also drawn in~\cite{Helmboldt:2014iya, Marko:2017yvl},
where the relation between pole and curvature mass has also been investigated.

A very good estimate for the pole mass
can be obtained from a Pad\'e approximant of the propagator
around the zero crossing of the real part of $\Gamma^{(2)}$, if the
pole is sufficiently close to the Minkowski axis.
This is certainly the case for the spectral
functions depicted in~\Fig{fig:LPApYspec}. Our result for the
different masses is shown together with the order parameter
$\langle \sigma \rangle$ in~\Fig{fig:mass_plot}.

While the difference is negligibly small for the pion at very
small temperatures, as expected since the momentum dependence
of the two-point correlator, shown in~\Fig{fig:euclMomDep}, is
very close to unity. A small deviation can be seen around the
phase transition, where the momentum dependence becomes
maximal. The more interesting case is the sigma, as its
spectral function has no clear mass pole for low
temperatures. Across the phase transition the curvature mass
can be used as an order parameter, since it is directly
related to the chiral condensate, see e.g.\ \cite{Pawlowski:2014zaa}. 
The pole mass of the sigma
mesons behaves more gentle across the phase transition in
comparison to the curvature mass, but still
exhibits a clear minimum at the cross over. The larger
mismatch between the two masses can again be explained by the
significantly stronger momentum dependence of the sigma meson
at small and medium temperatures compared to the pion,
c.f.~\Fig{fig:euclMomDep}. In general the qualitative strength
of the mass difference can already be obtained from the
temperature dependence of the constant wave function
renormalizations, as $Z_k(0,|\vec{p}|) \approx Z_k(0,k)$.

Furthermore, one can use~\eq{eq:pole} in order to extract the pole mass from the two-point function. Combining the spectral representation~\eq{eq:spectral_euclidean} with~\eq{eq:pole} one obtains
\begin{align} \label{eq:laplace_mass}
G_E(t,\vec{p}) = \frac{1}{Z(0,0)} \int_0^\infty\!\!\! \mathrm{d}\eta\ e^{-\eta t} \rho(\eta,\vec{p})\;,
\end{align}
i.e. the Fourier transformed propagator, which reduces to a
calculation of a Laplace transform. In the case of a mass pole
we were able to extract the correct pole mass
from~\eq{eq:laplace_mass} again, an example for the pion at a
temperature $\unit[138]{MeV}$ is shown
in~\Fig{fig:FourierPlot}. \Eq{eq:laplace_mass} allows us to
calculate the contributions from different structures in the
spectral function individually, i.e. we find the expected
exponential decay~\eq{eq:pole} from the mass pole and
empirically the contribution from the Landau cut is very well
described by additionally introducing a quadratic time
dependence in the exponent. Furthermore, their sum is already
sufficient to describe the full time dependence at reasonable
times, for extremely large times the behaviour is trivially
dominated by the necessary numerical cut
$\eta_\text{\tiny{min}}$. Unfortunately we were unable to
extend this definition of the pole mass to the regime without
a pole mass in the spectral function, due to a combination of
a lack of the functional form of the Landau cut and numerical
uncertainties.
\subsection{Spectral functions at finite external
  momentum} \label{sec:finite_ext_mom} Since there is only a
very small difference between LPA and LPA'+Y in our current
settings, the results with finite external momentum are
calculated in LPA due to reduced numerical cost, due to which
we also refrained from extrapolating our results to
$\varepsilon\to 0$. The results are depicted
in~\Fig{fig:ExtMom} for various temperatures and external
momenta.
The main differences between spectral functions at vanishing and finite external momentum is the uniform Lorentz boost of the mass pole and the unitarity cut, as well as the transport peak, at frequencies $\omega< |\vec{p}\, |$,
arising from space-like scattering processes at finite temperature and momenta with the heat bath, a detailed discussions about the involved kinematics can be found in~\cite{Tripolt:2014wra}. Furthermore the transport peak is, like the Landau cut, not Lorentz invariant as it couples directly to the heat bath.

\subsubsection*{Lorentz invariance}
A non-trivial consistency check of our results at finite external momenta is obtained by looking at the Lorentz invariance at a vanishing (or small) temperature. A Lorentz invariant function must only depend on $\omega^2-\vec{p}^{\, 2}$, i.e. for the spectral function this translates to the property that $\rho(\sqrt{\omega^2+\vec{p}^{\, 2}},\vec{p})$ must be independent of $\vec{p}$. Our results for the spectral functions at finite external momenta are depicted for this momentum configuration in~\Fig{fig:lorentzInvariance} at a small temperature. The most notably breaking of Lorentz invariance is introduced by the finite value of the small parameter $\varepsilon$, but we also do not expect invariance for these parts of the spectral function. Additional breaking is visible for small frequencies in the sigma spectral function and for frequencies around the mass pole in the pion spectral function due to the small temperature present, c.f.~the discussion in~\Sec{sec:spectral_result} and~\Sec{sec:finite_ext_mom}. The remaining parts of the spectral functions, i.e. the position of the pole, the thresholds and the continuum part, show perfect Lorentz invariance. 

This is in contrast to most previous studies of spectral functions within the FRG where a regulator of the form~\eq{eq:LitimReg} was used, and therefore Lorentz invariance explicitly broken~\cite{Kamikado:2013sia, Tripolt:2014wra}, and demonstrates the strength of our approach.

\section{Conclusions}
In this work we put forward a direct calculation of finite
temperature spectral functions with the FRG in the $O(N)$
model. This direct computation is based on a
$O(4)$/Lorentz-invariant regularisation scheme, and can be
performed on a fully numerical level. i.e.\ including the full
momentum- and frequency dependence of correlation
functions. It demonstrates the applicability of the formalism
put forward in~\cite{Pawlowski:2015mia} at finite
temperatures.

The spectral functions for the pion and sigma meson are shown across
the phase transition as a function of frequency and momentum. The four
expected structures in the spectral functions, i.e. the mass pole,
unitarity cut, Landau cut and transport peak, are present and
discussed in detail.

The spectra obtained allowed us to investigate the important relation
between the curvature and the pole mass. We found that the
definition of the pole mass as inverse temporal length is accessible
within our framework, but unsuitable if particles are unstable and a
pole is unidentifiable in the spectrum. An analytic relation between
the pole and curvature mass was derived in the case of a stable
particle and qualitatively verified that the difference is driven by
non trivial momentum dependencies.

Furthermore, we have explicitly verified the Lorentz
invariance of the spectral function. Another major advantage
of the employed framework is its numerical accessibility,
which makes it easily usable in more complex theories. We hope
to report on applications to Yang-Mills theory and QCD in the
near future.

\acknowledgments We thank Anton~K.~Cyrol, Stefan Floerchinger,
Fabian Rennecke and Alexander Rothkopf for discussions.  This work is supported by EMMI, the
grant ERC-AdG-290623, the DFG
through grant STR 1462/1-1, the BMBF grant 05P12VHCTG, and is part of
and supported by the DFG Collaborative Re- search Centre ”SFB 1225
(ISOQUANT)”. It is also supported in part by the Office of Nuclear
Physics in the US Department of Energy’s Office of Science under
Contract No. DE-AC02-05CH11231.

\begin{appendices}
\appendix

\section{Regulator \& cut-off scale}\label{a:reg}

In order to use a Lorentz invariant regulator $R_k(p^2)$, we need to
deal with additional poles/branch-cuts necessarily introduced by the
non-trivial analytic structure of $R_k(p^2)$. While the propagator
$1/\Gamma_k^{(2)}$ is restricted to have poles on the
Minkowski axis, a Lorentz invariant regulator with non-trivial
momentum dependence that leaves all non-analyticities of the regulated
propagator $1/(\Gamma_k^{(2)}(q)+R_k(q))$ on the Minkowski axis is to
date not known. As mentioned in section~\Sec{sec:formalism} we use the
regulator introduced in~\cite{Pawlowski:2015mia}, i.e.
\begin{equation} 
R_k(p^2) = Z_\sigma \left[p^2 + \Delta m_r^2 \right] r_k\left(\frac{q^2+\Delta m_r^2}{k^2}\right)\; ,
\end{equation} \label{eq:4dreg}
for the shape function we chose the exponential one
\begin{equation}
r(x) = \frac{x^{m-1}}{e^{x^m}-1}
\end{equation}
with $m=2$. The mass-like term $\Delta m_r^2$ has the effect of pushing regulator poles away from the Euclidean axis, while only having a small impact on the analytic structure of the unregulated propagator. It is parametrised as a smooth theta function
\begin{equation}
\Delta m_r^2 = \alpha \frac{p_{0,\text{max}^2}}{1+\left(\frac{\beta k}{p_{0,\text{max}}}\right)^n}\;,
\end{equation}
where we have chosen our parameters according to~\Tab{tab:regParams}.
\begin{table}[b]
	\centering
	\begin{tabular}{|l || c | c | c | c|}
		\hline
		Parameter & $p_{0,\text{max}}$ & $\alpha$ & $\beta$ & $n$ \\ [0.5ex] 
		\hline
		Value & \unit[3.45]{GeV} & 2.5 & 0.44 & 150 \\ [1ex] 
		\hline
	\end{tabular}
	\caption{Values for the parametrisation of $\Delta m_r^2$.}
	\label{tab:regParams}
\end{table}
These parameters are chosen such that the initial
conditions, $\Gamma_\Lambda$, are unchanged compared
to the standard case $\Delta m_r^2=0$. In the vacuum
it is possible to show for LPA that the regulator
poles do not contribute to a certain frequency range
if $p_{0,\text{max}}$ is sufficiently large and the
available frequency range corresponds roughly to its
value. For finite temperature this is in not true
anymore, as the correction factor for simple poles at
position $z_0$ is proportional to
$n_B(z_0+\imag p_0)-n_B(z_0)$, which is only
sufficiently suppressed for $|z_0|/T \gg 1$. Therefore
we have restricted our frequency range to
$\unit[720]{MeV}$, where we have checked for explicit
independence of the results on the parametrisation of
$\Delta m_r^2$.

The large value of $\Delta m_r^2$, which enables us to
resolve a large range of frequencies, comes with the
downside that the cut-off scale $\Lambda_{\text{UV}}$
must be sufficiently large. In order to still fulfil
the requirement of unchanged initial conditions
compared to a vanishing $\Delta m_r^2$ we have chosen
$\Lambda_{\text{UV}} = \unit[8.28]{GeV}$, therefore
the interpretation of our results as an effective
theory of low energy QCD is strictly speaking not
possible. Nevertheless our results demonstrate the
applicability of the method to extract real time
correlation functions from the FRG via analytic
continuation and the qualitative features stay
unchanged compared to the usual O(N)-model with a
lower cut-off. 
We are unable to fix all
values to their physical ones, as we observe a loss of
solution similar to~\cite{Marko:2016wtw}, restricting
us from tuning to arbitrary IR values. Moreover we
also observed a smaller and smaller range of initial
values that do not lead to a break down of the
numerics as we increase our truncation. Therefore we
chose $f_\pi/m_\pi = 0.93$ and $m_\sigma/m_\pi = 2.09$,
which fulfils the requirement $m_\sigma/m_\pi > 2$,
resulting in the sigma being an unstable particle.
For all truncations the initial values where chosen such that
the curvature masses agree.
\section{Different regulators}
\label{a:RegCompare}
\begin{figure}
\includegraphics[width=\linewidth]{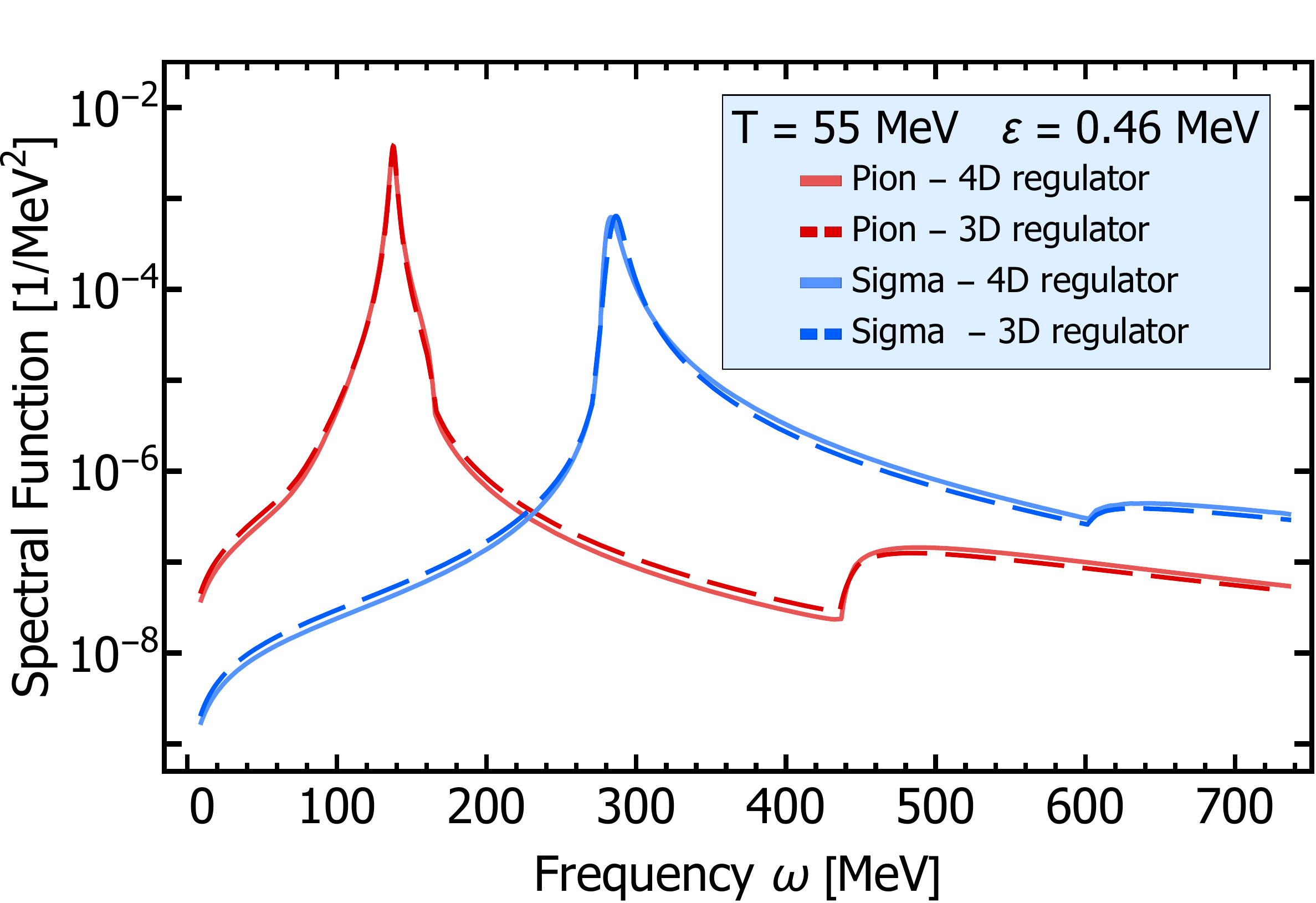}
\caption{Comparison of different regulators.}
\label{fig:RegCompare}
\end{figure}

In addition to the regulator described in~\App{a:reg}
it is also insightful to compare to the Lorentz
invariance breaking regulator~\eq{eq:LitimReg}, which
has been used in most FRG related works so far
\begin{equation} \label{eq:LitimReg}
R_k(\vec{p}^{\,2})=\left(k^2-\vec{p}^{\,2}\right) \Theta\left(k^2-\vec{p}^{\,2}\right)\; .
\end{equation}
This regulator has the advantage that the Matsubara
sum and the following analytic continuation can be
performed analytically. Therefore we compare our
results obtained with the Lorentz invariant
regulator~\eq{eq:4dreg} against the results obtained
with~\eq{eq:LitimReg} for the LPA case and vanishing
external momenta in~\Fig{fig:RegCompare}.
This shows that the difference between the two
regulators is marginal, which can be explained by the
small breaking of Lorentz invariance for $\vec{p}=0$,
c.f. the discussion in~\Sec{sec:finite_ext_mom}, and
the already very similar running of the Euclidean
system for the two regulators~\cite{PSSW} in this
regime. For larger temperatures the difference is
dominantly driven by the difference in the condensate,
leading to a general mismatch between the spectral
function.
\section{Numerical details} \label{a:numerics} We
followed the general workflow outlined
in~\cite{Cyrol:2016tym}. The flow equations are derive
using DoFun~\cite{Huber:2011qr}, traced and optimized
using the FormTracer~\cite{Cyrol:2016zqb}, utilizing
FORM~\cite{Kuipers:2012rf}, and solved using the
\textit{frgsolver}, a c++ framework developed and
maintained by the
fQCD-collaboration~\cite{fQCD:2016-10}.

In order to implement the procedure outlined
in~\Sec{sec:formalism}, the equations are solved as
\begin{multline}
\Gamma^{(2)}(\omega,\vec{p}) = \Gamma^{(2)}_\Lambda(\omega,\vec{p}) + \\
\lim_{\varepsilon \to 0} \int_{\Lambda}^{k_\text{IR}}\!\!\! \mathrm{d}k \int_0^\infty \!\!\! \mathrm{d}^3\vec{q}\   
\bigg\{ \sum_{q_0=2 \pi n T} + \text{ Correction} \bigg\} \\ 
\text{Flow}\left[\Gamma^{(2)}_k(q_0,\vec{q},p_0=-\imag (\omega + \imag \varepsilon),\vec{p}) \right]
\end{multline}
where "Correction" term refers to the correction of
the occupation numbers arising implicitly from the
Matsubara sum. From the study of the flow equation
using a 3D regulator it is well
known~\cite{Pawlowski:2015mia, Jung:2016yxl} that the
equations reduce to delta functions for the imaginary
part when the limit $\varepsilon \to 0$ is taken by
means of the Sokhotski-Plemelj theorem. Due to our
Lorentz invariant regulator~\eq{eq:4dreg} we are
however not able to resolve the Matsubara sum
analytically and must take the limit numerically,
therefore greatly increasing the numerical cost of the
calculation as the delta functions (and derivatives
thereof) have to be resolved numerically for very
small values of $\varepsilon$. In practice the limit
is performed using Richardson extrapolation with
several small values of $\varepsilon$, the
independence of the result for different sets of
$\varepsilon$'s has been checked explicitly. From the
symmetries of the two-point correlator it is known
that the leading term in $\varepsilon$ of the
imaginary part behaves as $\mathcal{O}(\varepsilon^1)$
and the real part behaves as
$\mathcal{O}(\varepsilon^2)$, the exponents in the
extrapolation are chosen accordingly.

The correction term can in principle also take branch
cuts into account, as described
in~\Sec{sec:formalism}, and therefore allows for a
self consistent treatment. With the current methods, this is  
numerically insurmountable as it introduces a second
competing limit for the branch cut integrals.
\end{appendices}
\bibliography{../../bib_master}
\end{document}